\documentclass[12,draftclsnofoot,onecolumn]{IEEEtran} 

\usepackage{graphicx}
\DeclareGraphicsExtensions{.eps,.ps,.eps.gz,.ps.gz,.eps.Z}
\usepackage{float}
\usepackage{amsmath}
\usepackage{amssymb}
\usepackage{cite}
\usepackage{epstopdf}

\ifCLASSINFOpdf
\else
\fi

\begin{document}
\title{\huge{Another look in the Analysis of Cooperative Spectrum Sensing over Nakagami-$m$ Fading Channels}}
%
%
%

\author{Debasish~Bera,~\IEEEmembership{Student Member,~IEEE,}
        Sant S.~Pathak,~\IEEEmembership{Senior~Member,~IEEE,}
				I.~Chakrabarti,~\IEEEmembership{Member,~IEEE}
        and~George~K.~Karagiannidis,~\IEEEmembership{Fellow,~IEEE}
\thanks{Part of this work was presented at IEEE 77th Vehicular Technology Conf. (VTC Spring), Dresden, Germany, June, 2013.}
\thanks{D. Bera is with the G. S. Sanyal School of Telecommunications, Indian Institute of Technology (IIT), Kharagpur,
WB, 721302 INDIA (e-mail: debasish.bera@gssst.iitkgp.ernet.in). S. S. Pathak, and I. Chakrabarti are with Dept. of E and ECE, IIT Kharagpur, (e-mail: {ssp, indrajit}@ece.iitkgp.ernet.in).}
\thanks{George K. Karagiannidis is with Aristotle University of Thessaloniki, Greece and with Khalifa University, Abu Dhabi, UAE (e-mail: geokarag@auth.gr).}
}

\maketitle

\begin{abstract}
Modeling and analysis of cooperative spectrum sensing is an important aspect in cognitive radio systems. In this paper, the problem of energy detection (ED) of an unknown signal over Nakagami-$m$ fading is revisited.
Specifically, an analytical expression for the local probability of detection is derived, while using the approach of ED at the individual secondary user (SU), a new fusion rule, based on the likelihood ratio test, is presented. The channels between the primary user to SUs and SUs to fusion center are considered to be independent Nakagami-$m$. The proposed fusion rule uses the channel statistics, instead of the instantaneous channel state information, and is based on the Neyman-Pearson criteria. Closed-form solutions for the system-level probability of detection and probability of false alarm are also derived. Furthermore, a closed-form expression for the optimal number of cooperative SUs, needed to minimize the total error rate, is presented. The usefulness of factor graph and sum-product-algorithm models for computing likelihoods, is also discussed to highlight its advantage, in terms of computational cost. The performance of the proposed schemes have been evaluated both by analysis and simulations. Results show that the proposed rules perform well over a wide range of the signal-to-noise ratio.
\end{abstract}


%
\IEEEpeerreviewmaketitle

\section{Introduction}
\IEEEPARstart{I}{t} is well-known that most of the licensed spectrum is not fully utilized all the time \cite{FCC2003}, when fixed spectrum allocation is used. Moreover, the rapid deployment of new wireless devices and applications with growing data rates creates a spectrum scarcity problem. Cognitive radio networks (CRN) \cite{Haykin2005} is an emerging solution to the problem of inefficient use of allocated licensed spectrum.
In this approach, the secondary users (SUs) or cognitive radios (CR)s are allowed to sense the spectrum dynamically, identifing the spectrum holes i.e. in the absence of a primary user (PU) - in the target spectrum pool and opportunistically utilize it. 
\subsection{Motivation and Literature}
Spectrum sensing is the first critical step of the CR cycle \cite{Haykin2005} in order to dynamically utilize the unused spectrum. Sensing techniques can be classified as, a) \textit{Local Sensing:} Each SU individually and/or independently detects spectrum holes. Although this kind of sensing is sensitive to fading, shadowing, and model uncertainty, it has a simple implementation. A brief survey of different spectrum sensing techniques was presented in \cite{Ying-Chang2011}. Furthermore, it is shown in \cite{Sahai2004} that the energy detection (ED) is optimal for detecting zero-mean constellation signals, if no prior knowledge about PU's signal is available at the SU, except of the received signal power. Note, that ED is also popular due to the simplicity of its implementation \cite{Yucek2009}. 
b) \textit{Cooperative Sensing:} Information from multiple SUs are jointly used to detect spectrum holes, and to mitigate multipath, shadowing etc., by exploiting the spatial diversity among CRs. It enhances accuracy, reliability, and performance of sensing at the cost of complexity. Moreover, cooperative sensing is most effective, when collaborating CRs observe independent fading or shadowing \cite{Mishra2006,Yucek2009,YCLiang2008,Bera2012,YulongZou2012,FanLeiDuongElkashlan2014,YeohElkashlanKaragiannidis2015}. 
\par
Cooperative sensing may be further viewed as distributed detection problem, with the central coordinator to be the fusion center (FC). This is also known as centralized cooperative spectrum sensing (CCSS) and is investigated in this paper. A detailed survey on the distributed detection was presented in \cite{Viswanathan1997}. In this kind of detection, likelihood ratio test (LRT) rule is known to be optimal. However, a global optimal solution with coupled local best rules is known to be NP-hard and not available in closed-form \cite{Varshney1997}. 
Moreover, it was proved that ED is optimal for single-sensor detection \cite{PennaTWC2012}, while identical decision rule is asymptotically optimal for global decision in a large network \cite{Varshney1997}. 
LRT is implemented using either the \textit{Neyman-Pearson} (N-P) criterion (maximization of probability of detection subject to a constraint on probability of false alarm) or the \textit{Bayes} criterion (minimization of error) \cite{VanTrees1968}. 
\par 
Well-known sub-optimal fusion rules for ideal SU-FC (reporting) channels are: AND, OR, and VOTING \cite{Viswanathan1997}. Other sub-optimal decision fusion rules over noisy channels are Chair-Varshney fusion for the high signal-to-noise ratio (SNR), equal gain combining (EGC) for low SNR, and maximum ratio combining (MRC) for medium SNR \cite{Chen2004}. Recently, suboptimal fusion rules, known as optimal linear cooperation strategy for additive white Gaussian noise (AWGN) and linear-quadratic (LQ) strategy for ideal reporting channels have been discussed in \cite{QuanSayed2008} and \cite{Unnikrishnan2008}, respectively. Furthermore, in \cite{PennaTWC2012,Zarrin2008}, the authors used a probabilistic graphical approach to model the optimal LRT based fusion for cooperative spectrum sensing. Other sensing algorithms for PU detection, such as optimum matched filtering \cite{JunMa2009}, eigenvalue based detection \cite{Kortun2011}, cyclostationary feature detection \cite{Lunden2009}, generalized likelihood ratio test (GLRT) based sensing \cite{Zhang2010,Wang2010}, were also reported in literature. 
\par 
Good message-passing algorithms, like Pearl's belief-propagation (BP) algorithm and sum-product algorithm (SPA) \cite{Kschischang2001} over suitable graphical models, have been successfully employed for solving inference problems in various aplications, e.g. data mining, computational biology, statistical signal processing, and wireless communications. This approach provides exact solutions for acyclic graphs, while exhibiting a low computational complexity, compared to explicit methods \cite{Wymeersch2007}. 
Moreover, BP/SPA is inherently suitable for distributed implementation \cite{Kschischang2001}. Therefore, it becomes a practical and powerful tool to solve distributed inference problems, such as cooperative spectrum sensing in CRNs \cite{Zarrin2008,PennaTWC2012}.
\par
In the spectrum sensing literature, previous studies assume approximate channel statistics \cite{Unnikrishnan2008,Zeng2009,Kortun2011} or known \cite{QuanSayed2008,YCLiang2008,Bera2012} or estimated \cite{Zhang2010,Wang2010}, instantaneous channel state information (CSI). The effects of different signal models with known CSI on sensing have also been investigated in \cite{Atapattu2014,YCLiang2008}. Furthermore, the problem of energy detection of an unknown signal over Nakagami-$m$ fading was addressed in a few papers \cite{Digham2007,Herath2011}, but, the results were presented only for high SNRs. Regarding the decision fusion most of the works consider non-ideal sensing channels with ideal \cite{Viswanathan1997,ZhangMallik2009} or binary symmetric (BSC)/AWGN reporting channels \cite{QuanSayed2008,Zarrin2008,Bera2013}. Optimization of the CCSS scheme over Rayleigh fading and ideal reporting channels, was also addressed in \cite{ZhangMallik2009}. 
Furthermore, decision fusion over non-ideal reporting channels was introduced by \cite{Chen2004}, in the context of wireless sensor networks (WSN). However, multipath fading on sensing and reporting channels is common in a CRN and limits the performance of CCSS. However, none of the above works consider multipath fading on both PU-SUs and SUs-FC links, simultaneously. 
\par
Nakagami-$m$, is a general fading model \cite{Nakagami1960}, which often gives the best fit for land and indoor mobile applications \cite{SimonAlouini2004,SuraweeraGeorge2008,CZhong2011,DengElkashlanDuong2015}. However, cooperative spectrum sensing, in the presence of Nakagami-$m$ fading with channel statistics, is relatively less investigated. Moreover, SUs may be mobile in many applications, like object tracking, environment, habitat management etc., where the channel estimation is costly. Therefore, spectrum sensing over Nakagami-$m$ fading for a wide range of SNR and LRT based decision fusion without knowledge of instantaneous CSI, is useful for the system design. Moreover, in a large CRN it also involves conditional and unconditional independence on large number of random variables (RVs) and thus it leads to an increase of the overall system complexity. Hence, inference over graph with message passing is a good approach for this problem \cite{PennaTWC2012,Zarrin2008}.
\subsection{Contribution}
In this work, we study the performance of CCSS systems, by assuming that both PU-SU and SU-FC channels are Nakagami-$m$ and independent accross the SUs. 
The LRT statistics is computed through message passing over the representative NFG, in order to reduce the computation complexity. Specifically, the main contributions of this paper are as follows.
\begin{itemize}
	\item Derivation of an LR based fusion rule without knowledge of the instantaneous CSI. Closed-form expressions for the system level probabilities of detection, $P_D$, miss, $P_M$, and false alarm, $P_F$, are also derived. Furthermore, we present an alternate expression for the local probability of energy detection over Nakagami-$m$ fading. 
\end{itemize}
\begin{itemize}
	\item Determination of the optimal number of cooperating SUs, needed to minimize the total error rate, $P_{TOT}\!=\!P_M+P_F$, as a function of the SNR and the total number of SUs.
\end{itemize}
\begin{itemize}
	\item Modeling of CCSS using NFG and SPA in order to analyse the computation time complexity, compared with explicit method.
\end{itemize}
\subsection{Structure}
The rest of this paper is organized as follows. Section II refers to NFG and SPA, while Section III represents the system model, the assumptions used, and the problem formulation. Expressions for local probalilities of detection and false alarm are presented in Section IV, while the LRT-based fusion rule with NFG-SPA based model, closed-form analysis of system-level performance metrics and optimization of the CRN, are presented in Section V. Simulation results are reported in Section VI, and the complexity analysis and advantages of NFG-SPA settings are discussed in Section VII. Finally, Section VIII concludes the paper and propose some future research directions. 
\subsection{Notations}
Throughout this paper, $K$ denotes the total number of SUs present in the CRN, $N$ denotes the total number of (complex) signal samples available for detection, also known as time-bandwidth product. 
We use $a^{K}_{1}$ to denote the set of RVs $\left\{a_1,...,a_K\right\}$. Here, $E[.]$ denotes statistical expectation, $|a|$ denotes modulus of $a$, $P_a\left(.\right)$ denotes probability density function (\textit{pdf}) and $F_a\left(.\right)$ denotes cumulative distribution function (\textit{cdf}) of $a$. ${Nak}(m,.)$, $\mathcal{CN}(.,.)$, and $\mathcal{N}(.,.)$ denote Nakagami-$m$ distribution with fading severity parameter $m$, complex Gaussian, and real Gaussian distribution, respectively. 
\section{Factor Graph, Sum-Product Algorithm (SPA)}
Probabilistic graphical model (PGM) \cite{Bishop2006} is an effective way to represent the probabilistic dependencies between RVs. Well-known graphical models are Bayesian network (BN), Markov random field (MRF), Tanner graph (TG), junction tree (JT), and factor graph (FG) \cite{Bishop2006}. 
Among those, FGs are more general, since any BN, MRF or TG can be transformed as FG, with no increase in its representation size \cite{Kschischang2001}. Throughout the present work, we consider a version of FG, called normal factor graph (NFG) \cite{Wymeersch2007}, as the PGM. The primary goal of FG-SPA based modeling of CCSS is to reduce computational complexity. 
\subsection{Factor Graph}
Factor graph is a standard bipartite graphical representation of a mathematical relation between variables and local functions. There are two types of factor graphs \cite{Wymeersch2007}: conventional and normal (Forney-style) factor graph (NFG). In an NFG, functions or factors $\left\{f_j\right\}$ are represented by nodes and variables $\left\{x_l\right\}$ are represented by edges.\\
\textit{Example:} Consider a joint probability mass (density) function $f(.)$ of $L$ variables as $f\left(x_1,x_2,x_3,x_4,...,x_L\right)$.
Suppose, the function is factorized as 

\small
\begin{equation}\label{eq1}
f(x_1,x_2,...,x_L)=\frac{1}{Z}\prod^{J}_{j=1}f_j(s_j), s_j\subseteq \left\{x_1,x_2,...,x_L\right\},
\end{equation}
\normalsize
where $Z$ is a normalization factor. Alternatively, it can be represented through a graph with function nodes and variable edges. We consider the factorization with $L\!=\!7$ and $J\!=\!6$, where one variable is involved in more than two factors. Then,

\vspace{-.2cm}
\small
\begin{equation}\label{eq2}
 \begin{split}
f\left(x_1,x_2,x_3,x_4,x_5,x_6,x_7\right)\! &=\! \frac{1}{Z}f_A(x_1,x_2,x_3,x_4)f_B(x_1,x_5)\times \\
                            &\quad f_C(x_2,x_7)f_D(x_4)f_E(x_5)f_F(x_5,x_6).
 \end{split}
\end{equation}
\normalsize
\vspace{-0.4cm}
\begin{figure}[ht]
\centering
\includegraphics[scale=0.5]{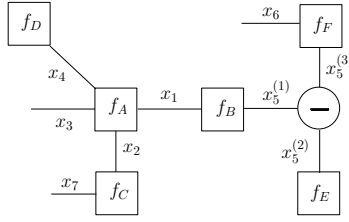}
\caption{Normal factor graph corresponding to Eq. (2).}
\label{Fig1}
\end{figure}

Fig.\ref{Fig1} depicts an example of \eqref{eq2} as a normal factor graph. We summarize the construction of NFG as follows:
\begin{itemize}
	\item Equality node, $\Theta$, indicates variables corresponding to more than two functions i.e. the node with degree, $D>2$.
\end{itemize}
\begin{itemize}
	\item Computation of marginal can be performed in an efficient and automated way by using SPA on factor graph.
\end{itemize}
\begin{itemize}
	\item A function can have many factorizations; therefore, it can have many factor graphs. As long as the graphs have no cycles, the same marginal will be computed for all.
\end{itemize}
	
\subsection{Sum-Product Algorithm and Message Passing}
Sum-product algorithm (SPA), also known as message-passing or belief-propagation (BP) algorithm, 
can often be applied successfully in situations, where exact solutions to the \textit{marginalize product-of-function} (MPF) problems become computationally intensive \cite{Kschischang2001,Bishop2006}. SPA operates over an NFG associated with the global function and computes various marginal probabilities by approximating through beliefs. Let us define the message from function node $f_j\in \mathcal{N}(x_l)$ to variable edge $x_l$ as $M_{f_j\rightarrow x_l}(x_l)$. The message from variable edge $x_l\in \mathcal{N}(f_j)$ to function node $f_j$ is denoted by $M_{x_l\rightarrow f_j}(x_l)$, where $\mathcal{N}(x_l)$ and $\mathcal{N}(f_j)$ are the set of neighboring functions of $x_l$ and the set of variables involved in function $f_j$, respectively. Message from edge $x_l$ to node $f_j$ is computed as 

\small
\begin{equation}\label{eq3}
 M_{x_l\rightarrow f_j}(x_l) \propto \prod^{}_{f\in \mathcal{N}(x_l)\text{\textbackslash} \left\{f_j\right\}}\!\!\!\!\!\!M_{f\rightarrow x_l}(x_l).
\end{equation}
\normalsize
and message from node $f_j$ to edge $x_l$ is computed as 
\small
\begin{equation}\label{eq4}
	M_{f_j\rightarrow x_l}(x_l)\!\propto\!\!\sum^{}_{X_j\in \mathcal{N}(f_j)\text{\textbackslash} \left\{x_l\right\}}\!\!\!\!\!\!\!\!\!\!\!\!f_j({X_j}) \!\!\!\! \prod^{}_{x_j\in \mathcal{N}(f_j)\text{\textbackslash} \left\{x_l\right\}}\!\!\!\!\!\!\!\!\!\!\!\!M_{x_j\rightarrow f_j}(x_j),
\end{equation}
\normalsize
where $\mathcal{N}(i) \text{\textbackslash} \left\{a\right\}$ denotes all the nodes/edges that are neighbors of edge/node $i$ except for node/edge $a$. In SPA, \textit{sum} is due to summation and \textit{product} is due to product operation in \eqref{eq4}. In case of continuous variables, summation operation is replaced by integration. The proportionality sign in \eqref{eq3} and \eqref{eq4} is used to indicate a normalization factor, such that the distribution sums/integrates to one. Final marginal for any variable $x_l$ is calculated as belief i.e. the product of all incomming mesages as 

\small
\begin{equation*}
b(x_l)\propto\prod^{}_{f\in \mathcal{N}(x_l)}M_{f\rightarrow x_l}(x_l).
\end{equation*}
\normalsize
\section{System Model for CCSS over Fading Channels} 
\begin{figure}[ht]
\centering
\includegraphics[scale=0.5]{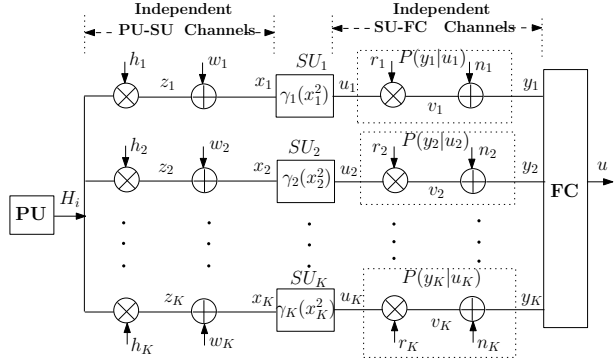} 
\caption{System block diagram of Centralized Cooperative Spectrum Sensing (CCSS) scheme with K SUs, one PU and one FC.}
\label{Fig2}
\end{figure}

The block diagram of CCSS system is shown in Fig. \ref{Fig2}. It consists of one PU, $K$ number of SUs, and one FC. All SUs are independently sensing the PU and then sending their local decisions to FC. Final decision is taken by the FC. 
\subsection{Assumptions}
Throughout this paper, Nakagami-$m$ fading is used to model rapid fluctuations of the amplitudes of a radio signal, and 
it is assumed that the average sensing duration is much shorter than the average \textit{busy-to-idle} and \textit{idle-to-busy} state transition periods of PU \cite{ZhangMallik2009}. Transmit power of PU is assumed to remain constant over a typical sensing period and \textit{a prior} probability of PU's traffic is unavailable at each SU.
\par
Furthermore, it is assumed that all SUs stay silent during the sensing interval, such that the spectral power remaining in the targeted band is transmitted only by the PU. Next, it is considered that all SUs use same transmit power relative to the PU (as in the interweave \textit{Dynamic Spectrum Access} (DSA) model \cite{Song2012}) and each SU makes a binary local decision (\textit{hard-sensing}) using ED \cite{Mishra2006}. At FC, the final decision ($u$) is taken when local decisions from all SUs are arrived. We formulate our problem by assuming that all \text{sensing} and \text{reporting} channels are time-invariant (during the sensing process), frequency-flat fading and statistically independent, across different SUs. 

\subsection{Problem Formulation}
Suppose that all SUs monitor the same frequency with the PU (Fig. \ref{Fig2}). 
Spectrum sensing at the $k$-th SU can be formulated as a binary hypothesis testing problem \cite{VanTrees1968}. The received signal samples at $SU_k$ for the two hypotheses can be modeled as 

\vspace{-0.15cm}
\small
\begin{align}\label{eq5}
&H_0: \text{PU is idle:}\ x_k(n) = w_k(n) \nonumber\\ 
&H_1: \text{PU is busy:}\ x_k(n) = z_k(n) + w_k(n),
\end{align}
\normalsize
where $n=1,...,N$, $k=1,...,K$, $w_k(n)\sim \mathcal{CN}(0,2\sigma^{2}_{w_k})$ is the sample of AWGN, and $z_k(n)$ represents the signal sample, received from the primary user if active. The signal is modeled as RV with average power of $E[|z_k(n)|^2]\!=\!\Omega_{z_k}$, which includes the channel gain. In practice, $z_k(n)$ and $w_k(n)$ are independent.
\par 
Classically, the received primary signal samples at $SU_k$ are assumed (reasonable approximation for unknown PU signals over fading; \cite{Zhang2010,Wang2010,PennaTWC2012}) to be independent and identically distributed (\textit{i.i.d.}) complex Gaussian RVs with zero mean and variance, $E[|z_k(n)|^2]$. This implies that the magnitude of the complex envelope 
is Rayleigh distributed. However, in practice, the received signal at each $SU_k$ is composed of a large number of resolved multipath components. 
Therefore, the magnitude of the envelope is the norm of an $m$-dimensional 
complex vector, where $m$ is the fading parameter of the Nakagami-$m$ distribution \cite{Beaulieu2005}. 
Therefore, $|z_k(n)|\sim {Nak}(m,\sigma_{z_k})$, and the average power of the received primary signal can also be written as $E[|z_k(n)|^2]\!\!=\!\Omega_{z_k}\!\!=\!2m\sigma^{2}_{z_k}$ \cite{Beaulieu2005}. The average received SNR of the $PU\!-\!SU_k$ link, measured at $SU_k$, is defined as 

\small
\begin{equation}\label{eq6}
\bar{\rho}^{(m)}_{{ps}_k}=\frac{m\sigma^{2}_{z_k}}{\sigma^{2}_{w_k}}. 
\end{equation}
\normalsize
\par
At $SU_k$, the ED computes the energy of the received signal over $N$ samples. The computed energy is compared with the threshold $\tau_k$, which is determined from a given local probability of false alarm, and the binary local decision $u_k\!\in\!\!\left\{-1,1\right\}$ is generated, where $u_k\!=\!-1$ and $u_k\!=\!1$ denote absence and presence of the PU, respectively. Therefore, the test statistic at $SU_k$ becomes 

\small
\begin{equation}\label{eq6}
t_k\!=\!\sum^{N}_{n=1}|x_k(n)|^{2}\gtrless \tau_k. 
\end{equation}
\normalsize
Each decision, $u_k$, is transmitted to the FC over an independent, frequency-flat fading channel. The signal received at the FC from $SU_k$ is 

\small
\begin{equation}\label{eq7}
y_k = v_k+ n_k, 
\end{equation}
\normalsize
where $n_k \sim \mathcal{CN}(0,2\sigma^{2}_{n_k})$ is the observation noise and $v_k$ is the secondary signal, over the $SU_k\!-\!FC$ link. Similarly, we assume $|v_k|\!\sim\! {Nak}(m,\sigma_{v_k})$, where the average power of the signal is $E[|v_k|^2]\!=\!\Omega_{v_k}\!=\!2m\sigma^{2}_{v_k}$. In practice, $v_k$ and $n_k$ are independent and the noise samples at FC and SUs are also independent across different $PU\!-\!SU_k\!-\!FC$ links. The average received SNR for $SU_k\!-\!FC$ link is defined as $\bar{\rho}^{(m)}_{{sf}_k}\!=\!\!\frac{m\sigma^{2}_{v_k}}{\sigma^{2}_{n_k}}$. 
\par
The vector of received signal at the FC from all SUs is denoted by $y^{K}_{1}=\left\{y_1,y_2,...,y_K\right\}$. For any $y^{K}_{1}$ the binary hypothesis problem at FC is

\small
\begin{align}
&I_0:\ \text{Primary user is idle:} &P(y^{K}_{1}|H_0) \nonumber\\
&I_1:\ \text{Primary user is busy:} &P(y^{K}_{1}|H_1),
\end{align}
\normalsize
where $P(y^{K}_{1}|I_0)$ and $P(y^{K}_{1}|I_1)$ are the distributions of $y^{K}_{1}$ in absence ($I_0$) and presence ($I_1$) of the PU, respectively. Final decision ($u$) is derived at the FC by LRT using these two distributions.
\par
We assume that SUs use the spectrum, whenever they detect a spectral hole (white space). The constraint in the system model is the probability of erroneous decision about the presence of the PU. Hence, for efficient utilization of spectrum, according to \textit{Neyman-Pearson} criteria \cite{VanTrees1968}, the system designer needs to minimize the probability of miss or maximize the probability of detection, subject to the constraint that the probability of false alarm satisfies a minimum requirement. 

\section{Local Performance Analysis}
This section presents the local performance analysis (probabilities of detection and false alarm at $SU_k$) for the system model defined in section-III.B. Throughout the analysis, the variable $a_k(n)$ is used as $a_k$ for notational simplicity. 
\subsection{Analysis Under the Hypothesis $H_0$}
Under $H_0$, the received signal contains only noise and thus, $x_k\sim \mathcal{CN}(0,2\sigma^{2}_{w_k})$. As the number $N$ of complex samples is considered as sensing period, the \textit{pdf} of $t_k$ under $H_0$ follows Chi-square distribution with $2N$ degrees of freedom \cite{Abramowitz1972}. In this case, the probability of false alarm at $SU_k$ is obtained as \cite{Digham2007} 

\small
\begin{align}\label{eq8}
p_{f_k}\!=\!\!\int^{}_{t_k>\tau_k}\!\!\!\!\!\!\!\!\!\!\!\!P(t_k|H_0) dt_k=\! \frac{\Gamma(N,\frac{\tau_k}{2\sigma^{2}_{w_k}})}{\Gamma(N)}
=\!1-\!\frac{\gamma(N,\frac{\tau_k}{2\sigma^{2}_{w_k}})}{\Gamma(N)},
\end{align}
\normalsize
where $\gamma(N,x)$, $\Gamma(N,x)\!=\!\Gamma(N)-\gamma(N,x)$, and $\Gamma(N)$ are the lower incomplete gamma, upper incomplete gamma, and complete gamma functions, respectively \cite{Abramowitz1972}. Assuming that the noise variance is perfectly known, the local threshold $\tau_k$ is obtained for a given $p_{f_k}$ as 

\small
\begin{align}\label{eq9}
\tau_k \!=\!2\sigma^{2}_{w_k}\Gamma^{-1}\left(N, \left(p_{f_k}\Gamma\left(N\right)\right)\right). 
\end{align}
\normalsize
%
\subsection{Analysis Under the Hypothesis $H_1$} 
Under the hypothesis $H_1$, the received signal contains both PU signal and noise. Therefore, the distribution of the test statistic depends on distribution of the envelope of the signal (i.e. $|x_k|$), which further depends on $|z_k|$. The unconditional \textit{pdf} of ${|x_k|}$ under $H_1$ is obtained using Bayesian approach, i.e. marginalizing the conditional \textit{pdf} of $|x_k|$ over $|z_k|$. Here, $|z_k|$ is Nakagami-$m$ distributed \cite{Nakagami1960} and is defined as 

\vspace{-0.2cm}
\small
\begin{align}\label{eq10}
P_{|z_k|}(c_k)
=\frac{2 {c_k}^{2m-1}}{\Gamma(m)} {\left(\!\frac{1}{2\sigma^{2}_{z_k}}\!\right)}^m \!\! e^{-\frac{{c_k}^2}{2\sigma^{2}_{z_k}}}, 
\end{align}
\normalsize
The \textit{pdf} of $|x_k|$ under $H_1$ is obtained as 

\small
\begin{align}\label{eq11}
&P_{|x_k|}(a_k|H_1) \!\!=\!\!\! \int^{\infty}_{0}\!\!\!\!\!P_{|x_k|\big{|}|z_k|}(a_k|c_k;H_1)P_{|z_k|}(c_k) dc_k \nonumber \\ 
=& \frac{{\left(\frac{\sigma^{2}_{w_k}}{A_{w_k}}\right)}^{m}e^{-\frac{{a_k}^2}{2\sigma^{2}_{w_k}}}}{\Gamma(m) \sqrt{2\pi \sigma^{2}_{w_k}}}\!\!\left[\Gamma(m) {}_1\!F_1\left(m,\frac{1}{2},{a_k}^2 D_{w_k}\right)+ \right. \nonumber \\ 
& \left. \frac{a_k\sqrt{2}\Gamma(m+\frac{1}{2}){}_1\!F_1\left(m+\frac{1}{2},\frac{3}{2},{a_k}^2 D_{w_k}\right)}{\frac{\sigma_{w_k}}{\sigma_{z_k}}\sqrt{A_{w_k}}}\right],
\end{align}
\normalsize
where $A_{w_k}\!=\!\left(\sigma^{2}_{z_k}\!+\!\sigma^{2}_{w_k}\right)$, $D_{w_k}\!=\!\frac{\sigma^{2}_{z_k}}{2\sigma^{2}_{w_k}A_{w_k}}\!=\!\frac{\alpha_{{ps}_k}}{2\sigma^{2}_{w_k}}$, $\alpha_{{ps}_k}\!=\!\frac{\sigma^{2}_{z_k}}{\sigma^{2}_{z_k}+\sigma^{2}_{w_k}}\!=\!\frac{\bar{\rho}^{(m)}_{{ps}_k}}{m+\bar{\rho}^{(m)}_{{ps}_k}}$, $\bar{\rho}^{(m)}_{{ps}_k}\!=\!\frac{m\sigma^{2}_{z_k}}{\sigma^{2}_{w_k}}$, and ${}_1\!F_1(.;.;.)$ is the confluent hyper-geometric function \cite[(13.1.2)]{Abramowitz1972}. 
\par
As, $t_k\!=\sum^{N}_{n=1}|x_k(n)|^{2}$ and $x_k(n)$ is complex, $t_k$ has $2N$ degrees of freedom. After a transformation of the variables in \eqref{eq11} we get the \textit{pdf} of $t_k$ under $H_1$ as

\vspace{-0.15cm}
\small
\begin{align}\label{eq12}
&P_{Nak}(t_k|H_1) = \frac{1}{2\sqrt{t_k}}[P_{|x|\big{|}H_1}(\sqrt{t_k})+P_{|x|\big{|}H_1}(-\sqrt{t_k})] \nonumber \\
=& \frac{{t_k}^{N}e^{-\frac{t_k}{2\sigma^{2}_{w_k}}}}{N! \left(2\sigma^{2}_{w_k}\right)^{(N+1)}}{\left(\!\frac{m}{m+\bar{\rho}^{(m)}_{{ps}_k}}\!\right)}^m \!\! {}_1F_1\left(m;N+1;t_k D_{w_k} \right).
\end{align}
\normalsize
Integrating \eqref{eq12} using Laguerre polynomials \cite{wolfram}, \cite[(6.9.2.36)]{Erdelyi1953}, and after some algebra \cite[(6.5.12)]{Abramowitz1972}, the local probability of detection can be written as 

\vspace{-0.15cm}
\small
\begin{align}\label{eq13}
&p^{(m)}_{d_k}=1-F_{Nak}(t_k|H_1)\nonumber \\
&=\!\!1\!-\!{\left(\!\frac{m}{m\!+\!\bar{\rho}^{(m)}_{{ps}_k}}\!\right)}^m\! \frac{\left(\!\frac{\tau_k}{2\sigma^{2}_{w_k}}\!\right)^{(\!N\!+\!1\!)}}{N! \ e^{\frac{\tau_k}{2\sigma^{2}_{w_k}}}} \Phi_2\left(\!m,1;N\!+\!1;\frac{\alpha_{{ps}_k}\tau_k}{2\sigma^{2}_{w_k}},\! \frac{\tau_k}{2\sigma^{2}_{w_k}} \!\right).
\end{align}
\normalsize
In \eqref{eq13}, $\Phi_2\left(.,.;.;.,.\right)$ is the hypergeometric function of two variables \cite{Abramowitz1972}. The steps for \eqref{eq12}-\eqref{eq13} are given in Appendix B. 
Note that Eq. \eqref{eq13} is more general (as it holds $\forall m\geq\frac{1}{2}$) than \cite[(7)]{Digham2007}, which is restricted to integer $m$. Moreover, Eqs. \eqref{eq13} and \cite[(7)]{Digham2007} can be implemented via the MATHEMATICA software, which requires truncation of infinite series and computation of error bounds \cite{SofotasiosKaragiannidis2014,Herath2011}. 
However, a general closed-form expression of $p^{(m)}_{d_k}$ is intractable, due to the presence of ${}_1\!F_1(m;.;.)$, $e^{-t}$, and $t^{N}$ in \eqref{eq12}, simultaneously \cite{Abramowitz1972}. Therefore, it is interesting to find $p^{(m)}_{d_k}$ for specific values of $m$.
\\
\\
\textit{\underline{Case-I: $m\!=\!1$}}
\\
For $m\!=\!1$, \eqref{eq11} is simplified as 

\small
\begin{align}\label{eq14}
P_{|x_k|}(a_k|H_1)\! 
=\!\frac{\sigma_{w_k}e^{\!-\frac{{a_k}^2}{2\sigma^{2}_{w_k}}}}{A_{w_k} \sqrt{2\pi}}\!+\! \frac{\sigma_{w_k}B_{w_k} a_k}{A_{w_k}} \ e^{\!-\frac{{a_k}^2}{2A_{w_k}}}\ Q(-B_{w_k} a_k),
\end{align}
\normalsize
where $B_{w_k}\!\!=\!\!\frac{\sigma_{z_k}}{\sigma_{w_k}\sqrt{A_{w_k}}}$ and $Q(a)$ is the well-known Gaussian $Q$-function \cite{Abramowitz1972}. For the derivation of \eqref{eq14} see Appendix C. Setting $m\!=\!1$ in \eqref{eq12} and using \cite[(6.5.12)]{Abramowitz1972} we obtain 

\small
\begin{align}\label{eq15}
P_{(m=1)}(t_k|H_1)\!
=\! \frac{\left(1-\alpha_{{ps}_k}\right)e^{-\frac{t_k\left(1-\alpha_{{ps}_k}\right)}{2\sigma^{2}_{w_k}}}}{\left(\alpha_{{ps}_k}\right)^{N}2\sigma^{2}_{w_k}\Gamma(N)} \gamma\left(N,\! \frac{t_k \alpha_{{ps}_k}}{2\sigma^{2}_{w_k}}\right),
\end{align}
\normalsize
where $\bar{\rho}^{(1)}_{{ps}_k}\!\!=\!\!\frac{\sigma^{2}_{z_k}}{\sigma^{2}_{w_k}}$. The \textit{cdf} is obtained by integrating \eqref{eq15} over $t_k$ using \cite[(1.2.2.1)]{Prudnikov1992}. Hence, the probability of detection at $SU_k$ can be written as 
\small
\begin{align}\label{eq16}
p^{(m=1)}_{d_k}\!\!&=\!\!\int^{}_{t_k>\tau_k}\!\!\!\!\!\!\!P_{(m=1)}(t_k|H_1) \ dt_k \!=\!1-\!\int^{\tau_k}_{0}\!\!\!\!\!\!\!P_{(m=1)}(t_k|H_1) \nonumber \\
&=\!1-\frac{\gamma\left(\! N,\! \frac{\tau_k}{2\sigma^{2}_{w_k}}\!\right)}{\Gamma(N)} \!+\! \frac{e^{-\frac{\tau_k}{2A_w}}}{\left(\alpha_{{ps}_k}\right)^{N}}\!\times\!\frac{\gamma\left(\! N,\! \frac{\tau_k \alpha_{{ps}_k}}{2\sigma^{2}_{w_k}}\!\right)}{\Gamma(N)}.
\end{align}
\normalsize
\textit{\underline{Case-II: $m\!=\!2$}}
\par
Similarly, for $m\!=\!2$, \eqref{eq12} can be written as 

\small
\begin{align}\label{eq17}
&P_{(m=2)}(t_k|H_1)\!=\!\frac{{t_k}^{N}e^{-\frac{t_k}{2\sigma^{2}_{w_k}}}{\left(\!\frac{2}{2+\bar{\rho}^{(2)}_{{ps}_k}}\!\right)^{2}}}{\left(N\right)!\left(2 \sigma^{2}_{w_k}\right)^{N+1}} {}_1F_1\left(2;N+1;t_k D_{w_k} \right), 
\end{align}
\normalsize
where $\bar{\rho}^{(2)}_{{ps}_k}\!\!=\!\!\frac{2\sigma^{2}_{z_k}}{\sigma^{2}_{w_k}}$. With the help of Appendix D, the probability of detection can be expressed as 

\small
\begin{align}\label{eq18}
&p^{(m=2)}_{d_k}\!\!=\!1-\!\int^{\tau_k}_{0}\!\!P_{(m=2)}(t_k|H_1) \ dt_k \nonumber \\
=&1\!-\!\!\frac{\gamma\left(\!\! N\!-\!1,\! \frac{\tau_k}{2\sigma^{2}_{w_k}}\!\!\right)}{\Gamma(N-1)} \! +\!\frac{e^{-\frac{\tau_k}{2A_w}}}{\left(\alpha_{{ps}_k}\right)^{N\!-\!1}} \frac{\gamma\left(\!\! N\!-\!1,\! \frac{\tau_k \alpha_{{ps}_k}}{2\sigma^{2}_{w_k}}\!\!\right)}{\Gamma(N\!-\!1)}\left(\!1\!+\!\frac{\tau_k}{2A_{w_k}}\!\right)\nonumber \\
& +\! \left(N\!-\!1\right)\left(1-\alpha_{{ps}_k}\!\right)\left[1-\!\frac{e^{-\frac{\tau_k}{2A_{w_k}}}}{\left(\alpha_{{ps}_k}\right)^{N}}\frac{\gamma\left(\!\! N,\! \frac{\tau_k \alpha_{{ps}_k}}{2\sigma^{2}_{w_k}}\!\!\right)}{\Gamma(N)}\right].
\end{align}
\normalsize
\par
Similarly, using other values of $m$ in \eqref{eq12} and integrating over $t_k$, corresponding $p_{d_k}$'s can be obtained. For example, an expression for $p_{d_k}$, when $m\!=\!\frac{1}{2}$, is presented in Appendix F. 
\par
In the next subsection, we present another approach for the determination of $p_{d_k}$'s, following the same assumptions as in \cite{Mallik2010,Yacoub2010}. For the sake of bravity and simplicity, we consider an approximate model, which is valid for moderate and high SNRs.
\subsection{Approximate Complex Representation of the Nakagami-$m$ Envelope}%
In \cite{Yacoub2010}, it is shown that the exact distributions of real and imaginary parts of the complex signal for Nakagami-$m$ envelope are non-Gaussian. As a special case, either the in-phase or quadrature signal may be assumed as zero-mean Gaussian, while the other part will be non-Gaussian \cite{Yacoub2010}. In \cite{Mallik2010}, the distributions of in-phase and quadrature parts of a signal having Nakagami-$m$ fading envelope, are derived 
for both $m\!>\!1$ and $\frac{1}{2}\!\leq\! m\!<\!1$. 
However, at high SNRs, the proposed model of \cite{Mallik2010} closely approximates the distributions of real and immaginary parts of signals with Rician \cite[(59)]{Mallik2010} for $m>1$ and Hoyt \cite[(61)]{Mallik2010}, for $\frac{1}{2}\!\leq\! m\!<\!1$ fading envelopes. These approximations are considered here for simplicity. However, for $m=1$, $z_k(n)$ can be assumed as $\mathcal{CN}\left(0,2\sigma^{2}_{z_k}\right)$ for all SNRs \cite{Zhang2010,Wang2010,PennaTWC2012,Zarrin2008}. Therefore, with the help of \cite{Mallik2010}, the complex signals over Nakagami-$m$ fading can be written as 

\small
\begin{align}\label{eq19}
&z_k(n)\sim \mathcal{CN}\left(0,\frac{\Omega_{z_k}\left(1+b\right)}{2}+\frac{\Omega_{z_k}\left(1-b\right)}{2}\right) \text{, $\frac{1}{2}\leq m<1$} \nonumber\\
&z_k(n)\sim \mathcal{CN}\left(0,2\sigma^{2}_{z_k}\right) \ \text{, $m=1$ } \nonumber\\
&z_k(n)\sim \mathcal{CN}\left(\mu_{I_{z_k}}+j\mu_{Q_{z_k}}, \Omega_{s_k}\right) \ \text{, $m>1$},
\end{align}
\normalsize
where $\Omega_{s_k}\!\!=\!\!\left(\frac{m-\sqrt{m^2-m}}{m}\right)\Omega_{z_k}\!\!=\!\!2\sigma^{2}_{z_k}\left(m-\sqrt{m^2-m}\right)$, $\mu_{I_{z_k}}\!\!=\!\!\left(\sqrt{\Omega_{z_k}d}\right)\text{cos}\left(\varphi\right)$ and $\mu_{Q_{z_k}}\!\!=\!\!\left(\sqrt{\Omega_{z_k}d}\right)\text{sin}\left(\varphi\right)$, $d\!\!=\!\!\sqrt{\frac{m-1}{m}}$, and $b\!\!=\!\!\sqrt{\frac{1-m}{m}}$. $\varphi$ is defined in \cite[(39)]{Mallik2010}. 
PU is active under $H_1$ and the distribution of the decision statistic is obtained from that of the signal. At high SNR, the distribution of $z_k(n)$ can be approximated according to \eqref{eq19} for different range of $m$, as in \cite{Mallik2010}. Thus, the associated probabilities of detection at $SU_k$ for different values of $m$ are obtained as follows 

\small
\begin{align}\label{eq20}
&p^{\left(\frac{1}{2}\leq m<1\right)}_{d_k}\!=\!1-\!\frac{\gamma\left(\!\!\frac{N}{2},\frac{\tau_k}{\left(\!\Omega_{z_k}\left(1+b\right)+2\sigma^{2}_{w_k}\!\right)}\!\!\right)}{\Gamma\left(\frac{N}{2}\right)}\!-\!\frac{\gamma\left(\!\!\frac{N}{2},\frac{\tau_k}{\left(\!\Omega_{z_k}\left(1-b\right)+2\sigma^{2}_{w_k}\!\right)}\!\!\right)}{\Gamma\left(\frac{N}{2}\right)} \nonumber \\
&p^{(m=1)}_{d_k}\!=1-\!\frac{\gamma\left(\!N,\frac{\tau_k}{2\left(\sigma^{2}_{z_k}+\sigma^{2}_{w_k}\right)}\!\right)}{\Gamma\left(N\right)} \nonumber \\
&p^{(m>1)}_{d_k}\!=Q_{N}\left(\sqrt{\mu_{z_k}},\sqrt{\frac{2\tau_k}{\Omega_{s_k}+2\sigma^{2}_{w_k}}}\right), 
\end{align}
\normalsize
where $Q_{N}\left(a,b\right)$ is the generalized Marcum-$Q$ function \cite{Abramowitz1972}. 
Here, the probability of missed detection is defined as $p^{(m)}_{m_k}\!=\!1-\!p^{(m)}_{d_k}$. For the derivation of \eqref{eq20} see Appendix G.
\subsection{Comparisons of the Two Models}
\begin{figure}[!t]
\centering
\includegraphics[scale=0.52]{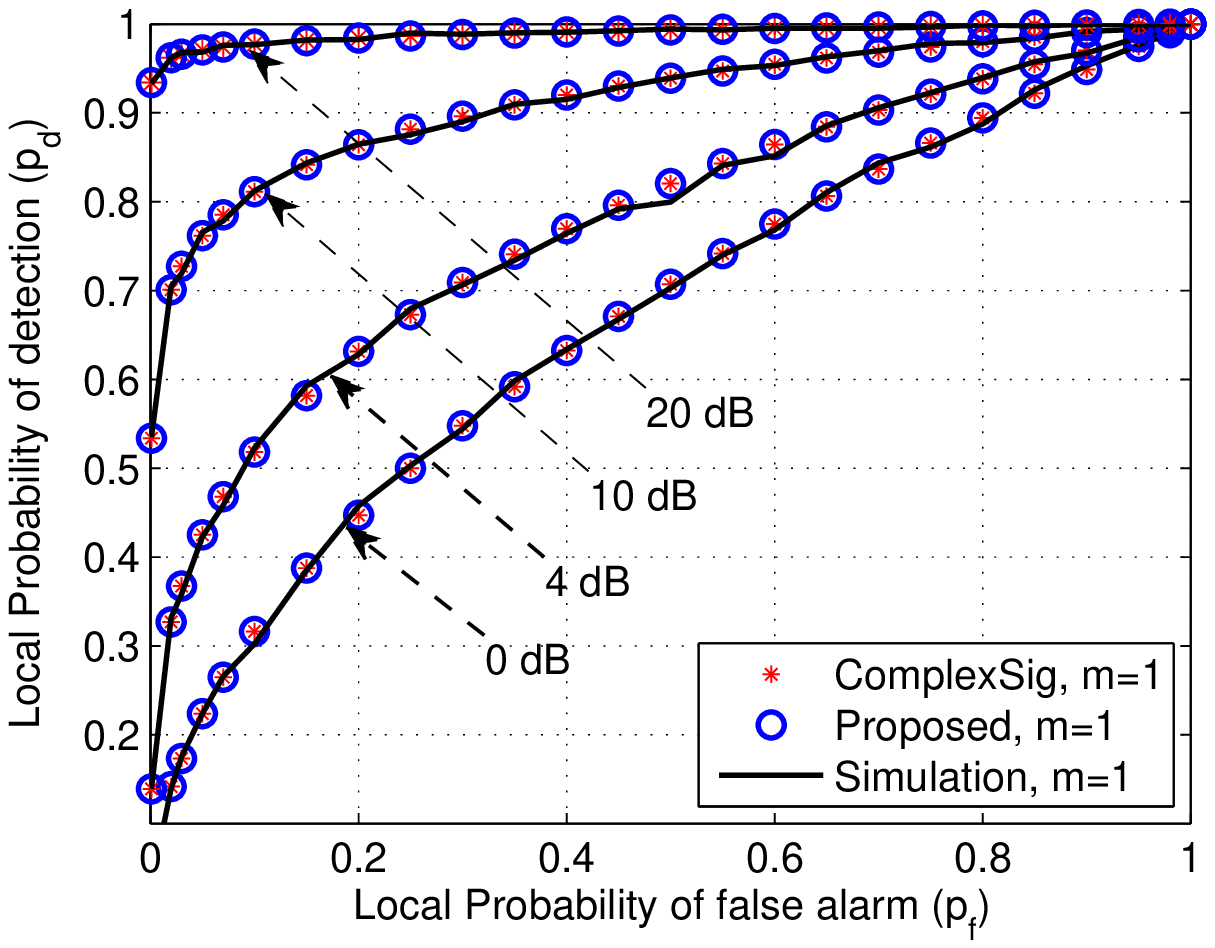} 
\vspace{-0.15cm}
\caption{ROCs over Nakagami-$m$ fading for different SNRs, with $N\!=\!10$ and $m\!=\!1$.}
\label{Fig3}
\end{figure}
\begin{figure}[!t]
\centering
\includegraphics[scale=0.52]{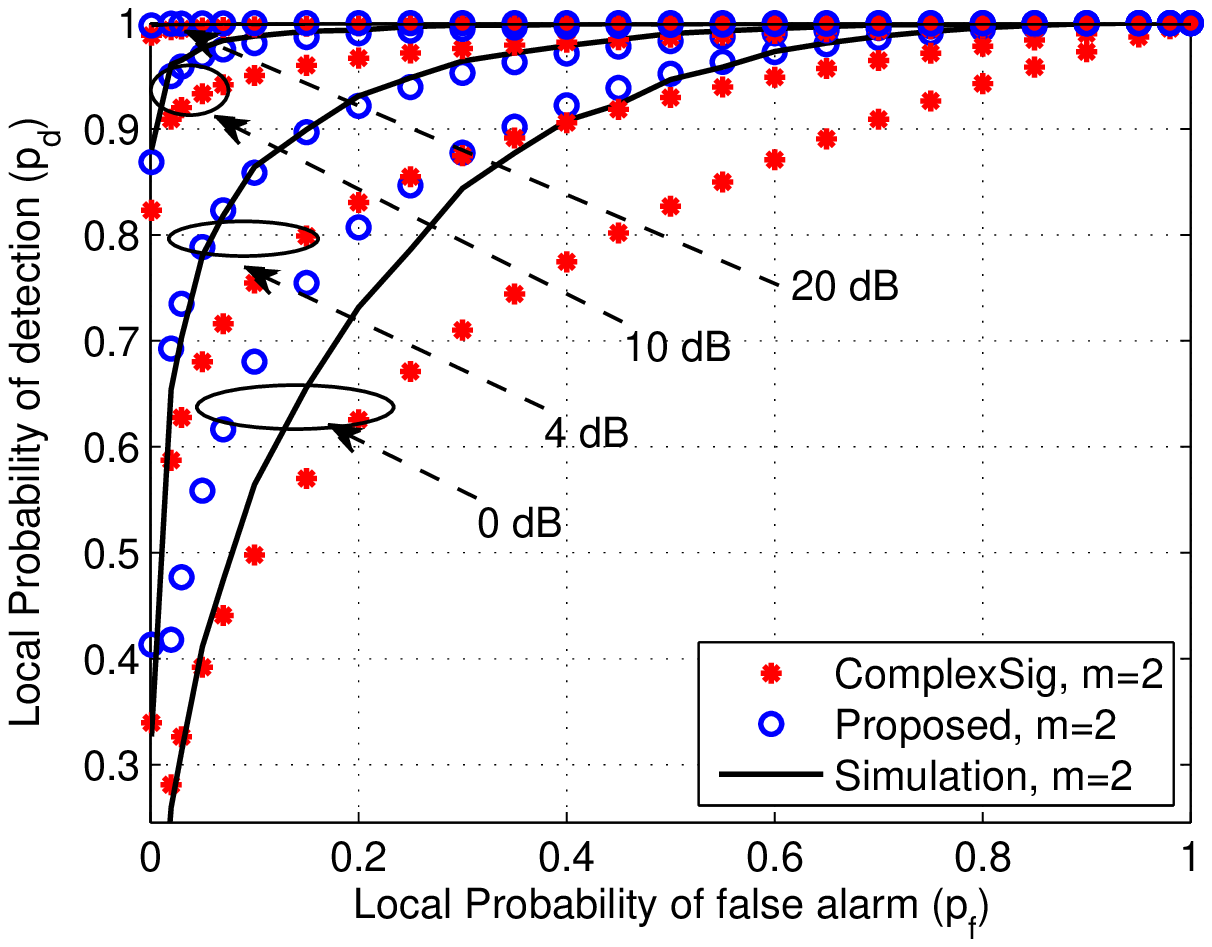}
\vspace{-0.15cm}
\caption{ROCs over Nakagami-$m$ fading for different SNRs, with $N\!=\!10$ and $m\!=\!2$.}
\label{Fig4}
\end{figure}
\begin{figure}[!t]
\centering
\includegraphics[scale=0.5]{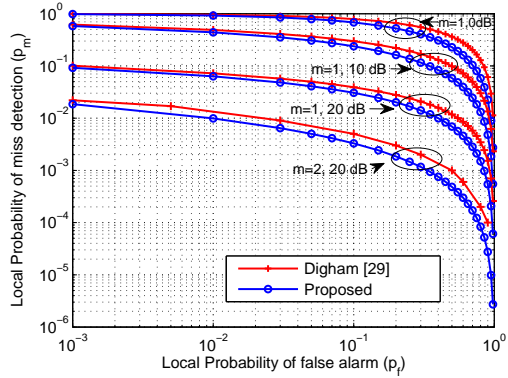}
\vspace{-0.15cm}
\caption{Comparisons of CROC curves over Nakagami-$m$ fading for different SNRs, with $N=10$ and $m=1,2$.}
\label{Fig5}
\end{figure}
Figs. \ref{Fig3} and \ref{Fig4} plot receiver operating characteristic (ROC), i.e. $p_{d_k}$ vs $p_{f_k}$ curves for different values of SNR $4$ dB, $10$ dB, and $20$ dB, over Nakagami-$m$ fading with $m=1$, and $2$, respectively. 
\par
It is observed from Fig. \ref{Fig3} that both analytical models for $m=1$ perfectly match with simulations for all SNRs. Further Fig. \ref{Fig4} shows that, for $m=2$, at very low SNR none of the models matches with simulations, but, it becomes closer to the proposed model in \eqref{eq18}. However, at moderate SNRs ($4$ dB to $10$ dB) the simulations match with the results of the proposed approaches, while it matches perfectly 
at high SNR ($20$ dB). Therefore, we can say that for $m>1$, the analysis based on the complex signal model \cite{Mallik2010} is only suitable for high SNRs. However, \eqref{eq20} is reasonable and also validates the well-known approximation of $z_k(n)$ for $m=1$ \cite{Zarrin2008,Zhang2010,Wang2010}. Note that, the analytical models proposed in this paper suit well over a wide range of SNR for $m\geq1$. A similar comparative study can be presented for $\frac{1}{2}\leq m<1$. It can also be proved that the proposed model in \eqref{eq57} suits better than the complex one in \eqref{eq20} over a wide range of SNR. In this paper, we drop the figure for $\frac{1}{2}\leq m<1$ due to space limitations. 
\par
Fig. \ref{Fig5} plots complementary ROC i.e. $p_{m_k}$ vs $p_{f_k}$ curves for different SNRs ($0$ dB, $10$ dB $20$ dB) over different fading channels ($m\!=\!1,2$) and compares the proposed model with Digham et. al's \cite{Digham2007}. It is observed that the model proposed in this paper is better than \cite{Digham2007} over a wide range of SNR. Hence, we always consider the analytical models of \eqref{eq16}, \eqref{eq18}, and \eqref{eq56} for further system-level analysis in the next section. 


\section{System-level Performance Analysis } 
\subsection{Probability Models of the Detection Problem}
\begin{figure}[!t]
\centering
\includegraphics[scale=0.6]{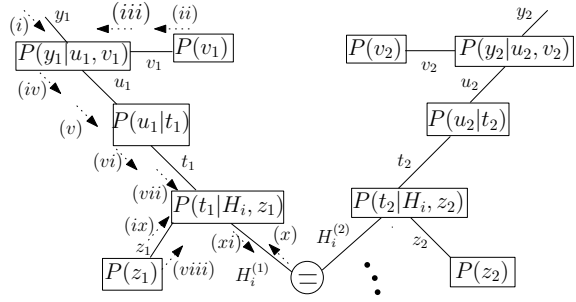} 
\caption{Forney-style factor graph for joint distribution of interest $P(z^{K}_{1}\!,t^{K}_{1}\!,u^{K}_{1}\!,v^{K}_{1}\!,y^{K}_{1}|H_i)$. The graph is shown for two PU-SU-FC channels.}
\label{Fig6}
\end{figure}

The CSS problem, described in Fig. \ref{Fig2}, can also be viewed as a system-on-graph (Fig. \ref{Fig6}). Thus, we can model the problem as inference over the representative NFG and try to solve it by passing messages over the graph using SPA. This approach can adopt all the unknown parameters, such as signals, channel effects, noise, and complex dependencies in a single framework. It is shown that the FG-SPA approach finds the desired likelihoods in an automated way. Accomplishable reduced complexity is shown in Section-VII. 
\par
As explained above, the goal is to find the likelihood functions, $P(y^{K}_{1}|H_0)$ and $P(y^{K}_{1}|H_1)$, in order to compute the LRT statistic and thus, to solve the distributed detection problem. These likelihoods can be obtained by marginalizing the joint probability distribution of interest over all unknown variables. Thus the detection problem is mapped to a Bayesian inference one, in order to find the likelihoods through message passing via SPA over the representative NFG. 
The joint probability distribution $P(H_i,z^{K}_{1},t^{K}_{1},u^{K}_{1},v^{K}_{1},y^{K}_{1})$ represents the CCSS model of Fig. \ref{Fig2}. Likelihood functions, represented by $P(y^{K}_{1}|H_i)$ for $i\!\in\!\left\{0,1\right\}$, can be evaluated as 

\vspace{-0.25cm}
\small
\begin{align}\label{eq21}
&P(y^{K}_{1}|H_i)\!=\!\!\! \int \!\!\!P(z^{K}_{1}\!,t^{K}_{1}\!,u^{K}_{1}\!,v^{K}_{1}\!,y^{K}_{1}\!|H_i)dv^{K}_{1} du^{K}_{1} dt^{K}_{1} dz^{K}_{1}, 
\end{align}
\normalsize
where $P(z^{K}_{1},t^{K}_{1},u^{K}_{1},v^{K}_{1},y^{K}_{1}|H_i)$ is the joint distribution of interest. This can be further factorized as

\vspace{-0.25cm}
\small
\begin{align}\label{eq22}
 &P(z^{K}_{1},t^{K}_{1},u^{K}_{1},v^{K}_{1},y^{K}_{1}|H_i) \nonumber\\
 &= P(y^{K}_{1}|u^{K}_{1}\!,v^{K}_{1})P(u^{K}_{1}|t^{K}_{1})P(t^{K}_{1}|H_i,z^{K}_{1})P(z^{K}_{1})P(v^{K}_{1}) \nonumber \\
 &=\!\prod^{K}_{k=1}\!P(y_k|u_k,v_k) P(u_k|t_k) P(t_k|H_i,z_k) P(z_k) P(v_k).
\end{align}
\normalsize
%
The last line in \eqref{eq22} holds because the channels are independent accross the SUs. Fig. \ref{Fig6} represents the NFG for the joint distribution of interest $P(z^{K}_{1},t^{K}_{1},u^{K}_{1},v^{K}_{1},y^{K}_{1}|H_i)$. Each branch represents the PU-SU-FC path of Fig. \ref{Fig2}. The equality node, $\Theta$, indicates that variable $H_i$ is associated with more than two functions i.e. with all $P(t_k|H_i,{z_k})$. It computes likelihoods from the joint distribution by employing \textit{hard decisions} at SUs. The graph has no cycle, as all PU-SU-FC channels are statistically independent. Therefore, SPA can compute the exact marginals over the graph. 
\subsection{Computation of Messages in NFG-SPA Settings}
In NFG, probability functions are represented by nodes and variables are represented by associated edges. The desired likelihoods $P(y^{K}_{1}|H_0)$ and $P(y^{K}_{1}|H_1)$ have to be computed from the graph. By applying the SPA as message computation rule \cite{Wymeersch2007}, intermediate messages are computed and passed between the nodes of the FG. The message propagation follows a single step of natural scheduling. As the graph has no cycle, the computation of messages starts from the half edge ($y_k$) (edge connected to only one node) and leaf node ($P({z_k})$ and $P({v_k})$) and proceed from node to node. The message ($M_{P({z_k})\rightarrow z_k}$) from the leaf node ($P({z_k})$) to the connecting edge ($z_k$) is the marginal value of the function (node) with respect to that variable (edge). For half edge, the message ($M_{y_k\rightarrow P(y_k|u_k)}$) from the edge to the node is initialized with the value $1$. For intermediate nodes, all incoming messages to that node are computed first and every message is computed only once. 
The messages are indexed with $(i),...,(xi)$ and are shown in Fig. \ref{Fig6} on the corresponding edges of the graph. Dotted arrows show the flow of messages for computing the marginals. Step-by-step computations are presented in Appendix A.

The marginal of $H_i$ on $k-$th branch is computed from the final messages of interest similarly as \cite[(21)]{Zarrin2008} 

\small
\begin{align}\label{eq24}
&g({H_{i}}^{(k)})= \int^{}_{t_k}\!\!\int^{}_{z_k}\!\!\!M_{P(t_k|H_{i},z_k)\rightarrow H_{i}} \times M_{H_{i} \rightarrow P(t_k|H_{i},z_k)} dz_k dt_k \nonumber \\
&=\! P(y_k|u_k\!=\!-1)\!\int^{\tau_k}_{0}\!\!\!\!\!\!\!\!P(t_k|H_{i})dt_k \!+\! P(y_k|u_k\!=\!1)\!\int^{\infty}_{\tau_k}\!\!\!\!\!\!\!\!P(t_k|H_{i})dt_k,
\end{align}
\normalsize
where 
\small
\begin{align}
P(t_k|H_{i})\!=\!\int^{}_{z_k}\!\!P(t_k|H_{i},{z_k})P({z_k})dz_k
\end{align}
\normalsize
is obtained by marginalizing over ${z_k}$, and $\tau_k$ is local threshold at $k$-th SU. 
\par 
The final message on each branch is computed as the product of messages from node $P(t_k|H_{i},{z_k})$ to edge $H_i$ and from edge $H_i$ to node $P(t_k|H_{i},{z_k})$. The desired likelihood functions for cooperative sensing are obtained from the marginal distribution of $H_i$ and computed as the accumulated final message (product over all branches) over all edges ${H_{i}}^{(1)},..,{H_{i}}^{(K)}$ for $i\!\in\!\left\{0,1\right\}$ from \eqref{eq24} as 

\small
\begin{equation}\label{eq25}
 P(y^{K}_{1}|H_i)= \prod^{K}_{k=1} g({H_{i}}^{(k)}). 
\end{equation}
\normalsize
Likelihoods are computed in automated way using NFG-SPA. 
\subsection{Analysis of Decision Fusion at FC}
LRT based decision fusion is performed at the FC. It requires likelihoods under $H_0$ and $H_1$ for each received signal $y_k$. Using \eqref{eq24}, the likelihood under $H_0$ can be viewed as the message under $H_0$ and can be written as \cite[(23)]{Zarrin2008}

\vspace{-0.15cm}
\small
\begin{align}\label{eq26}
P(y_k|H_0)\!&=\!M_{P(t_k|{z_k},H_0)\rightarrow H_0} \nonumber \\
&=\! P(y_k|u_k=-1)(1-P_{f_k})+P(y_k|u_k=1)P_{f_k}. 
\end{align}
\normalsize
Similarly, the likelihood under $H_1$ can be written as 

\vspace{-0.15cm}
\small
\begin{align}\label{eq27}
P(y_k|H_1)\!&=\! M_{P(t_k|z_k,H_1)\rightarrow H_1} \nonumber \\
&=\!P(y_k|u_k=-1)(1-p_{d_k})+P(y_k|u_k=1)p_{d_k}.
\end{align}
\normalsize
As the observations are independent, the LRT for choosing $H_1$ in N-P method can be written with help of \eqref{eq25} as

\vspace{-0.15cm}
\small
\begin{align}\label{eq28}
L(y^{K}_{1}\!) \!\!=\!\! \frac{P(y^{K}_{1}\!|H_1\!)}{P(y^{K}_{1}\!|H_0\!)} \!\! 
=\!\!\! \prod^{K}_{k=1}\!\!\frac{P(\!y_k|\!-\!1)(\!1\!-\!p_{d_k}\!)\!+\!P(\!y_k|1)p_{d_k}}{P(\!y_k|\!-\!1)(\!1\!-\! p_{f_k}\!)\!+\!P(\!y_k|1)p_{f_k}}\! \geq\! \lambda,
\end{align}
\normalsize
where $\lambda$ is the threshold at FC. In N-P settings, this is obtained by solving 

\small
\begin{align*}
\int^{}_{L(y^{K}_{1})>\lambda}\!\!P(y^{K}_{1}|H_0) \ dy^{K}_{1}=\int^{\infty}_{\lambda}\!P(L|H_0) \ dL\!=\!P_F,
\end{align*}
\normalsize
for a constraint on probability of false alarm at FC. 
\par
According to \eqref{eq7}, $n_k\! \sim \! \mathcal{CN}(0,2\sigma^{2}_{n_k})$ and the envelope of the received signal at FC over $SU_k-FC$ link is Nakagami-$m$ distributed i.e. $|v_k|\sim \! Nak(m,\sigma_{v_k})$. In general, $P(y_k|u_k)$ follows \eqref{eq11} by replacing $x_k$, $\sigma^{2}_{z_k}$ and $\sigma^{2}_{w_k}$ with $y_k$, $\sigma^{2}_{v_k}$ and $\sigma^{2}_{n_k}$, respectively. 
For $m\!=\!1$, as $P(y_k|u_k)$ follows \eqref{eq14}, the LRT statistic (using Appendix E) can be written as 

\small
\begin{align}\label{eq29}
L(y^{K}_{1})\!=\! \prod^{K}_{k=1}\frac{e^{\!-\frac{{y_k}^2}{2\sigma^{2}_{n_k}}}\!+\!\left\{p^{(m=1)}_{d_k}\!-Q\left(\!B_n y_k\!\right)\right\} \sqrt{2 \pi} B_{n_k} {y_k} e^{\!-\frac{{y_k}^2}{2A_{n_k}}}}{e^{\!-\frac{{y_k}^2}{2\sigma^{2}_{n_k}}}\!+\!\left\{p_{f_k}\!-Q\left(\!B_{n_k} y_k\!\right)\right\} \sqrt{2 \pi} B_{n_k} {y_k} e^{\!-\frac{{y_k}^2}{2A_{n_k}}}}, 
\end{align}
\normalsize
Similarly, for $m\!=\!2$, the LRT statistic (using Appendix E) is given by 

\small
\begin{align}\label{eq30}
L(y^{K}_{1}\!)\!\!=\!\!\prod^{K}_{k=1}\!\!\frac{B_{n_k} y_k e^{-\frac{{y_k}^2}{2\sigma^{2}_{n_k}}} \!+\!\left\{\!p^{(m=2)}_{d_k}\!\!-\!Q(\!B_{n_k} y_k\!)\!\right\}\! \sqrt{2\pi}R_{n_k} e^{-\frac{{y_k}^2}{2A_{n_k}}}}{B_{n_k} y_k e^{-\frac{{y_k}^2}{2\sigma^{2}_{n_k}}} \!+\!\left\{p_{f_k}\!-\!Q(\!B_{n_k} y_k\!)\right\}\! \sqrt{2\pi}R_{n_k} e^{-\frac{{y_k}^2}{2A_{n_k}}}}, 
\end{align}
\normalsize 
where $A_{n_k}\!\!=\!\!\left(\sigma^{2}_{v_k}+\sigma^{2}_{n_k}\right)\!\!=\!\!\left(1+\frac{\bar{\rho}^{(m)}_{{sf}_k}}{m}\right)\!\sigma^{2}_{n_k}$, $B_{n_k}\!\!=\!\!\frac{\sigma_{v_k}}{\sigma_{n_k}\sqrt{A_{n_k}}}\!\!=\!\!\frac{\sqrt{\alpha_{{sf}_k}}}{\sigma_{n_k}}$, $\bar{\rho}^{(m)}_{{sf}_k}\!\!=\!\!\frac{m\sigma^{2}_{v_k}}{\sigma^{2}_{n_k}}$, $\alpha_{{sf}_k}\!\!=\!\!\frac{\sigma^{2}_{v_k}}{\sigma^{2}_{v_k}+\sigma^{2}_{n_k}} \!\!=\!\!\frac{\bar{\rho}^{(m)}_{{sf}_k}}{m+\bar{\rho}^{(m)}_{{sf}_k}}$, $R_{n_k}\!\!=\!\!\left(2+ B^{2}_{n_k} {y_k}^{2}\right)$, and $p_{f_k}$ is obtained from \eqref{eq8}. The $p_{d_k}$'s are obtained from \eqref{eq16} and \eqref{eq18} for $m=1$ and $2$, respectively. Similarly, other likelihood ratios may be obtained by substituting corresponding values of $m$ in \eqref{eq28} with associated $p^{(m)}_{d_k}$'s. The LRT statistic for $m\!=\!\frac{1}{2}$ is derived in Appendix F. 
\par
According to this approach, local thresholds are derived, based on the probability of false alarm, and therefore, CSI is not needed 
at each SU. However, local probabilities of detection and final LRT statistics, i.e. $L(.)$ values, depend on channel statistics instead of instantaneous CSI. Moreover, system-level thresholds for LRT at FC are selected based on the $L(.)$ values. Therefore, we can state that, for the overall system design, the proposed fusion rule requires the knowledge of channel statistics, i.e. average SNRs $\bar{\rho}^{(m)}_{{ps}_k}$ and $\bar{\rho}^{(m)}_{{sf}_k}$ instead of the instantaneous CSI. 

\subsection{Closed-form Analysis}
Eqs. \eqref{eq29} and \eqref{eq30} are LR based fusion rules, but, the derivation of closed-form expressions for $P_D$ and $P_F$ from them, seems to be a hard task. However, LR based optimum fusion rule can be approximated as Chair-Varshney one \cite{Varshney1997}, under the assumption of high SNR and identical detectors (i.e. $p_{d_k}=p_d$, $p_{f_k}=p_f$ $\forall k$) \cite{Chen2004}. It is already assumed that $y_k$'s are statistically \textit{i.i.d.} for large number of SUs. To derive the closed-form expressions for $P_D$ and $P_F$, we also define as $K_1$ the number of SUs for which $y_k\geq0$ and $K-K_1$ the number of SUs for which $y_k<0$. Hence, the log-likelihood-ratio (LLR) from \eqref{eq28} can be written in terms of $K_1$ and $K$ 

\small
\begin{align}\label{eq31}
&log(L(y^{K}_{1}))\! 
=\!(\! K\!\!-\!K_1\!)Log\frac{P(y_k\! < \!0|-1)(\!1\!-\! p_d)\!+\!P(y_k\! < \!0|1)p_d}{P(y_k\! < \!0|-1)(\!1\!-\! p_f\!)\!+\!P(y_k\! < \!0|1)p_f} \nonumber \\ 
& + K_1 Log\frac{P(y_k \geq 0|-1)(1\!-\! p_d)+P(y_k \geq 0|1)p_d}{P(y_k \geq 0|-1)(1\!-\! p_f)+P(y_k \geq 0|1)p_f}.
\end{align}
\normalsize
Therefore, $K_1$ is binomial $(K, p)$ distributed, where the probability of success $p$ is defined as $p=P(y_k \geq 0)$. Let us denote, $p_1$ and $p_0$ are the probabilities of success under $H_1$ and $H_0$, respectively. Then, the system-level detection performance, i.e. $P_D$, $P_F$ and $P_M(=1-P_D)$ can be computed using the binomial distribution. Associated closed-form solutions are 

\small
\begin{equation}\label{eq32}
P_D = \sum^{K}_{j=K_1} \dbinom{K}{j} {p_1}^j (1-p_1)^{K-j},
\end{equation}
\normalsize
\small
\begin{equation}\label{eq33}
P_F = \sum^{K}_{j=K_1} \dbinom{K}{j} {p_0}^j (1-p_0)^{K-j},
\end{equation}
\normalsize
where $K_1$ takes values from $0$ to $K$. For $m=1$, $p_1$ and $p_0$ can be evaluated as (see Appendix E for the derivation)
%
%

\small
\begin{equation}\label{eq34}
  \left.
	\begin{aligned}
    &p_1 \!=\!\frac{1}{2}\!+\!\frac{(p_{d}-\frac{1}{2})\sigma_{v_k}}{\sqrt{\sigma^{2}_{v_k}\!+\!\sigma^{2}_{n_k}}}\!=\!\frac{1}{2}\!+\!\left(p^{(m=1)}_{d}-\frac{1}{2}\right) \sqrt{\alpha_{{sf}_k}}\quad\\ 
		&p_0 \!=\!\frac{1}{2}\!+\!\frac{(p_{f}-\frac{1}{2})\sigma_{v_k}}{\sqrt{\sigma^{2}_{v_k}\!+\!\sigma^{2}_{n_k}}}\!=\!\frac{1}{2}\!+\!\left(p_{f}-\frac{1}{2}\right) \sqrt{\alpha_{{sf}_k}}
  \end{aligned}
	\right\},
\end{equation}
\normalsize
Similarly, $p_1$ and $p_0$ are obtained for other values of $m$ with associated $p_{d}$ (see Appendix E and F). Hence, we state that N-P test is equivalent to J-out-of-K rule when
\begin{itemize}
	\item All PU-SU-FC channels are statistically independent.
\end{itemize}
\begin{itemize}
	\item All SU's are employing identical decision rules and transmitting hard local decisions to FC.
\end{itemize}
The theoretical ROC curve at FC is obtained from \eqref{eq32}-\eqref{eq33}. 

\subsection{Optimization of Cooperative Spectrum Sensing}
In general, when $P_D$ increases, $P_F$ also increases and as a result $P_M$ decreases. However, an effective system design should always minimize the total error rate, $P_{TOT}\!=\!P_M+P_F$. An optimal voting rule that minimizes Bayes risk function has been addressed in \cite[(pp. 94)]{Varshney1997}. Minimization of $P_{TOT}$ over ideal SU-FC channels using some linear approximation has also been reported in \cite{ZhangMallik2009}. Next, we investigate the optimal number, $l_{opt}$, of SUs required to minimize the total error rate over non-ideal SU-FC channels. 
\par
Let, $l$ be the number of SUs required to achieve certain values of $P_M$ and $P_F$. From \eqref{eq32} and \eqref{eq33}, let us define a function $D(l)=P_F-P_D$. Therefore, 

\vspace{-0.2cm}
\small
\begin{align}\label{eq35}
&D(l)=(P_M + P_F - 1) \nonumber \\
&=\! \sum^{K}_{j=l} \! \dbinom{K}{j} \left[{p_0}^j (1\!-\!p_0)^{K\!-j}-{p_1}^j (1\!-\!p_1)^{K\!-j}\right].  
\end{align}
\normalsize
\begin{figure}[!t]
\centering
\includegraphics[scale=0.5]{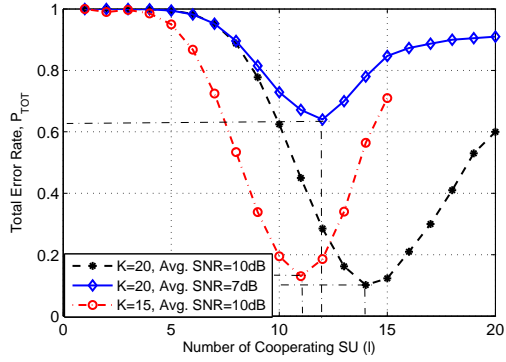}
\vspace{-0.2cm}
\caption{Number of cooperating SUs ($l$) vs $P_{TOT}$ curve when all channels as Nakagami-$1$ faded. $p_f\!=\!0.03$, $N\!=\!20$, $K\!=$ 20 and 15.}
\label{Fig7}
\end{figure}
From the properties of ROC curve \cite{VanTrees1968}, it is known that $p_d \geq p_f$. Then, from \eqref{eq34} and \eqref{eq54} for $m=1$ and $2$, respectively, holds that $p_1 \geq p_0$. Under this condition, it is evident from \eqref{eq35} that minimization of $P_{TOT}$ is equivalent to the minimization of $D(l)$. Now, when $l$ is increased by 1, it holds that 

\vspace{-0.3cm}
\small
\begin{align}\label{eq36}
D(l+1)\!-\!D(l)\!=\!\dbinom{K}{l}\! \left[{p_1}^l (1\!-\!p_1)^{K\!-l}\!-\!{p_0}^l (1\!-\!p_0)^{K\!-l}\right].  
\end{align}
\normalsize
%
From \eqref{eq36}, $D(l+1)\geq D(l)$ if,

\small
\begin{align}\label{eq37}
& \ {p_1}^l (1\!-\!p_1)^{K\!-l} \geq {p_0}^l (1\!-\!p_0)^{K\!-l}  \nonumber \\
\text{or,} \ &l \times ln\frac{p_1}{p_0} \geq (K-l) \times ln\frac{1-p_0}{1-p_1}   \nonumber \\
\text{or,} \ &l \geq \frac{K}{1+\beta} \ \text{, where}\ \beta \!=\! \frac{ln\frac{p_1}{p_0}}{ln \frac{1-p_0}{1-p_1}}.
\end{align}
\normalsize
Similarly, when $l$ is decreased by 1, 

\small
\begin{align}\label{eq38}
&D(l)-D(l-1)\nonumber \\
&=\dbinom{K}{l-1}\! \left[{p_1}^{l-1} (1\!-\!p_1)^{K\!-l+1}-{p_0}^{l-1} (1\!-\!p_0)^{K\!-l+1}\right].  
\end{align}
\normalsize
From \eqref{eq38}, $D(l)< D(l-1)$ if

\small
\begin{align}\label{eq39}
&{p_1}^{l-1} (1-p_1)^{K-l+1} < {p_0}^{l-1} (1-p_0)^{K-l+1}  \nonumber \\
\text{or,} \ &(l-1) \times ln\frac{p_1}{p_0} <  (K-l+1) \times ln\frac{1-p_0}{1-p_1}\nonumber \\
\text{or,} \ &l < 1+\frac{K}{1+\beta} \ \text{, where}\ \beta \!=\! \frac{ln\frac{p_1}{p_0}}{ln \frac{1-p_0}{1-p_1}}.
\end{align}
\normalsize
Thus, from \eqref{eq37} and \eqref{eq39} the optimal number of SUs, which minimizes $P_{TOT}$ can be written as 

\small
\begin{align}\label{eq40}
l_{opt} = \left\lceil \frac{K}{1+\beta}\right\rceil\, 
\end{align}
\normalsize
where 
\small
\begin{align*}
\beta \!=\! \frac{ln\frac{p_1}{p_0}}{ln \frac{1-p_0}{1-p_1}},
\end{align*}
\normalsize
and $p_1$, $p_0$ are given by \eqref{eq34} or \eqref{eq54}. Therefore, $l_{opt}$ depends on $K$, $p_0$, and $p_1$, subject to the constraint $p_1 \geq p_0$. These probabilities of success under $H_1$ and $H_0$ further depend on the average SNRs of SU-FC channels, local thresholds ($\tau_k$), noise variance ($\sigma^{2}_{w_k}$) of individual $SU_k$'s, and average SNRs of PU-SU channels. Hence, it is observed that the optimal number of SU ($l_{opt}$) depends on $K$, $\tau$, and average SNRs $\bar{\rho}_{ps}$ and $\bar{\rho}_{sf}$. 
Moreover, it is concluded that a fast spectrum sensing algorithm can be executed by considering only $l_{opt}$ SUs instead of all $K$ in a CRN, with statistically independent channels.
\par
Fig. \ref{Fig7} plots the number of cooperating SUs vs $P_{TOT}$ for different values of $K$ and different channel conditions. 
The figure shows that, for a fixed average SNR of $10$ dB over SU-FC channels, the optimal number of cooperating SUs increases from $11$ to $14$, while the total number of SUs increases from 15 to 20, respectively. It can also be observed that as $K$ increases, $P_{TOT}$ of corresponding $l_{opt}$ decreases. However, to achieve a given $P_{TOT}$ (say, 0.63) with fixed $K=20$, the required number of cooperating SUs ($l$) increases from 10 to 12, as the average SNR decreases from $10$ dB to $7$ dB. It infers that, for a fixed values of $K$, the value of $l$ increases to achieve a given (bounded) $P_{TOT}$, as SNR decreases. Hence, we can say that there exist an optimal number of cooperating SUs for different $K$ and different SNRs, subject to minimization of $P_{TOT}$.

\section{Simulations and Discussions}
In this section, simulation results are presented to evaluate the proposed CCSS scheme. The parameters used in the simulations are: 
$K=5$, and $10$, $N=20$ and number of Monte Carlo iterations $=\!10,000$. Independent, frequency-flat, Nakagami-$m$ fading with $m\!=\!1$ and $2$, and $\sigma^{2}_{w_k}\!\!=\!\!\sigma^{2}_{n_k}\!\!=\!1$, are considered. 
In the following, ED threshold ($\tau_k$) at $SU_k$, $\forall k$, are obtained from \eqref{eq9}, to maintain a target local probability of false alarm $p_{f_k}\!\!=\!\!0.03$. However, each $SU_k$ has different probabilities of detection ($p_{d_k}$) based on the different valuse of $m$ (e.g. \eqref{eq16}, \eqref{eq18} for $m=1,2$, respectively). The system-level performance is quantified by the ROC. The average SNRs of sensing and reporting channels are defined as $\bar{\rho}^{(m)}_{{ps}_k}(\text{dB})\!\!=\!\!10\!\log_{10}\!\frac{m\sigma^{2}_{z_k}}{\sigma^{2}_{w_k}}$ and $\bar{\rho}^{(m)}_{{sf}_k}(\text{dB})\!\!=\!\!10\!\log_{10}\!\frac{m\sigma^{2}_{v_k}}{\sigma^{2}_{n_k}}$, $\forall k$, respectively. 
\par
\begin{figure}[!t]
\centering
\vspace{-0.1cm}
\includegraphics[scale=0.493]{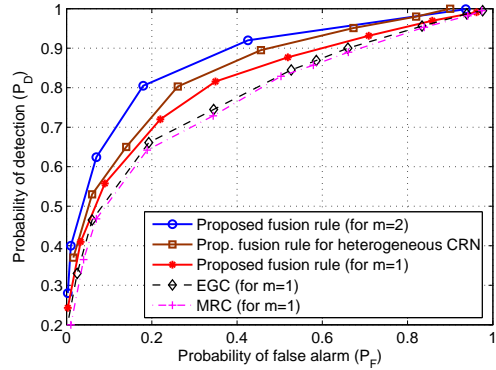}
\vspace{-0.2cm}
\caption{Comparison of ROC curves for various fusion rules over Nakagami-$1,2$ sensing and reporting channels where mean of $\left\{\bar{\rho}_{ps}\right\}\approx\!4$ dB and $\left\{\!\bar{\rho}_{sf}\!\right\}\approx\!4$ dB with $K\!=\!10$, $N\!=\!20$, $p_f\!=\!0.03$.}
\label{Fig9}
\end{figure}

Fig. \ref{Fig9} describes the ROC curves at different LRT threshold levels, for different fusion statistics of the CCSS. It is assumed that, $\left\{\bar{\rho}_{{ps}_{k}}\right\}^{K}_{k=1}\!\!\approx\!\left\{\!-4, -2, 0, 2, 3, 5, 10, 8, 7, 11\right\}$ dB and $\left\{\!\bar{\rho}_{{sf}_{k}}\!\right\}^{K}_{k=1}\!\!\!\approx\!\left\{\!-5, -3, -1, 0, 2, 4, 7, 12, 10, 14 \right\}$ dB, $m\!=1,2$, and $K\!=\!10$. It can be seen from Fig. \ref{Fig9} that the performance of the LRT based proposed fusion rule is better than the two suboptimal schemes, e.g. EGC and MRC \cite{Chen2004}, when $m=1$. Fig. \ref{Fig9} also depicts that the ROC curve improves as $m$ increases for all channels, since the effect of fading decreases.

\par
A more practical heterogeneous scenario with 10 SUs is also considered in Fig. \ref{Fig9}. It is assumed that in the first six PU-SU-FC channels, $m=1$ and for the rest four channels, $m=2$. 
It is shown that ROC performance improves for the heterogeneous case, when compared with that of $m=1$, because of the SU's spatial diversity (i.e. different fading severity parameters). 
\begin{figure}[!t]
\centering
\vspace{-0.1cm}
\includegraphics[scale=0.493]{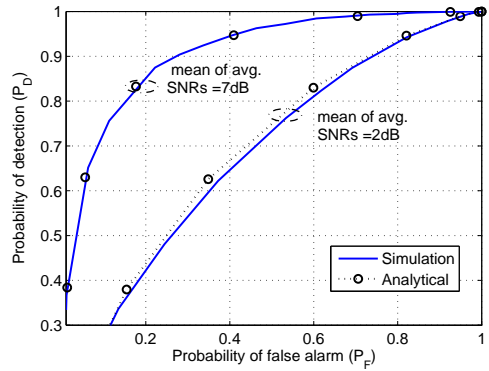}
\vspace{-0.2cm}
\caption{ROC curves over Nakagami-$1$ fading for different means of $\left\{\bar{\rho}_{sf}\right\}\approx\!{2}$ dB, $7$ dB with fixed mean of $\left\{\bar{\rho}_{ps}\right\}\approx \!4$ dB, $K\!=\!10$, and $p_f\!=\!0.03$.}
\label{Fig10}
\end{figure}
\par
Fig. \ref{Fig10} represents the ROC curves at FC of the LRT based proposed fusion rule, with $m=1$, $K=10$, and $p_f=0.03$. We consider single set of average SNRs for PU-SU links as $\left\{\bar{\rho}_{{ps}_{k}}\right\}^{K}_{k=1}\!\!\approx\!\left\{\!-4, -2, 0, 2, 3, 5, 10, 8, 7, 11\right\}$ dB, where the mean of all the average channel SNRs for the PU-SU links is $\approx 4$ dB. However, the two sets (low and high) of average SNRs for SU-FC links have been considered as $\left\{\!\bar{\rho}_{{sf}_{k_1}}\!\right\}^{K}_{k=1}\!\!\approx\!\left\{\!-10, -5, -2, -1, 0, 3, 5, 7, 10, 12\right\}$ dB and $\left\{\!\bar{\rho}_{{sf}_{k_2}}\!\right\}^{K}_{k=1}\!\!\!\approx\!\left\{\!-3,-1, 0, 4, 6, 8, 10, 12, 14, 20\right\}$ dB. Here, the means of all the average channel SNRs for SU-FC links are $\approx\!{2}$ dB and $7$ dB, respectively. The system-level analytical results and Monte Carlo simulations are obtained from \eqref{eq32}-\eqref{eq33} and \eqref{eq29}, respectively. As expected, ROC performance increases as average SNRs of SU-FC channels increase. 
\par
\begin{figure}[!t]
\centering
\vspace{-0.1cm}
\includegraphics[scale=0.51]{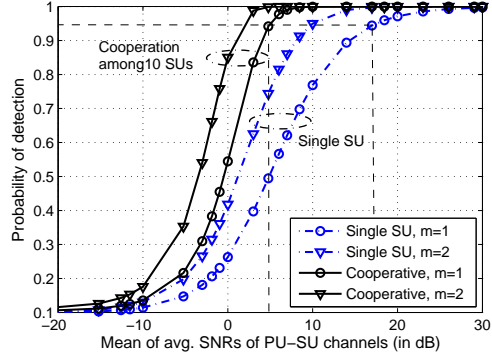}
\vspace{-0.2cm}
\caption{Probability of detection as a function of mean of avg. SNRs. Mean of $\left\{\bar{\rho}_{sf}\right\}\!\!\approx \!10$ dB and that of $\left\{\bar{\rho}_{ps}\right\}$ varies from $-20$ dB to $30$ dB. $K\!=\!10$, $N\!=\!20$, $p_f\!=0.03$, and $P_F\!=\!0.02$.}
\label{Fig11}
\end{figure}

For better clarity, in Fig. \ref{Fig11}, the local and system-level probabilities of detection are plotted as a function of the mean of the average SNRs of all the PU-SU channels. The cases of $m\!=\!1,2$ with $K\!=\!10$, $p_f\!=\!0.03$, and $P_F\!=\!0.02$, are considered. Here, SU-FC channels have mean of all the average SNRs $\approx \!10$ dB and that for PU-SU links are varying from $-20$ dB to $30$ dB. Note that there are significant variations for average PU-SU channel SNRs. In Fig. \ref{Fig11}, the performances of both cooperative and non-cooperative sensing schemes are presented for different $m$ values. It shows that cooperation among SUs significantly improves the probability of detection compared to the non-cooperative case, over a wide range of SNR. It is also observed that for $m\!=\!1$, the coopeative scheme achieves $0.95$ probability of detection at $5$ dB of SNR, whereas non-cooperative sensing reaches to the same at higher SNR ($17$ dB). Fig. \ref{Fig11} also shows that the probability of detection increases significantly for all SNRs, as $m$ increases from 1 to 2, in both the cases. 
\par
\begin{figure}[!t]
\centering
\vspace{-0.1cm}
\includegraphics[scale=0.54]{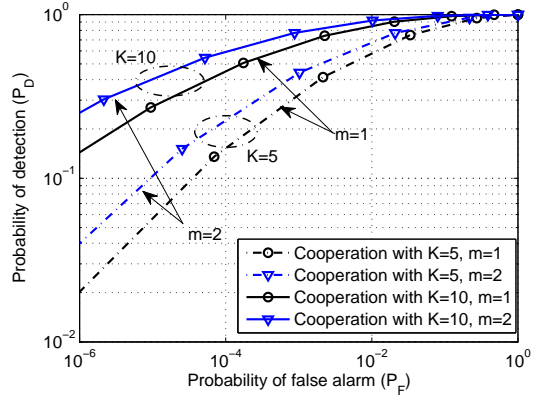} 
\vspace{-0.17cm}
\caption{ROC curves at FC for different $K(5,10)$ over Nakagami-$1,2$ channels with mean of $\left\{\bar{\rho}_{ps}\right\}\!\!\approx\!4$ dB, $\left\{\bar{\rho}_{sf}\right\}\!\!\approx\!9$ dB, $N\!=$20, $p_f\!=0.03$.}
\label{Fig12}
\end{figure}

Fig. \ref{Fig12} plots ROC curves obtained from \eqref{eq32}-\eqref{eq33} for different values of $K\!=\!5,10$ and $m\!=\!1,2$ with $p_f\!=\!0.03$. The case of $\left\{\bar{\rho}_{ps}\right\}\!\!\approx\!\left\{\!-4, -2, 0, 2, 3, 5, 10, 8, 7, 11\!\right\}$ dB and $\left\{\bar{\rho}_{sf}\right\}\!\!\approx\!\left\{1, 3, 5, 8, 9, 10, 13, 15, 17, 9,\right\}$ dB are considered. Here, the mean of the SNRs for PU-SU and SU-FC channels are $4$ dB and $9$ dB, respectively. For $K\!=\!5$, SNRs are drawn from index 3 to 7 of the above sets (to maintain the same mean for both $K$). It shows that detection performance increases significantly as the number of cooperating users increases from 5 to 10 for $m=1$ and $2$ both. The reason is the accumulation of information from more number of SUs. 
\section{Complexity Analysis and Advantage of NFG-SPA Based Approach for the Sensing Problem} 
The usefulness of the NFG-SPA based approach to cooperative spectrum sensing problem can be analyzed via the computational cost to find the marginal likelihoods. A function may be factorized in several ways, resulting in respective FGs. As long as the graphs are acyclic, the same marginals will be computed. The physical meaning of acyclic is that all PU-SU-FC channels are statistically independent. However, in correlated cases it may lead to more complex SPA, due to the presence of cycles in the graph \cite{Kschischang2001}. 
\par
The complexity of an algorithm is usually measured in terms of its performance as a function of the size of the input. Here, it depends on the number of functions (nodes) and variables (edges) \cite{Wymeersch2007}. 
Let us consider a function with $M$ variables that may be factorized (acyclic) in $F$ factors. We assume that each variable is defined over the same $X$ discrete-sampled values from a continuous domain or domain $X$ for discrete RVs. The function is represented by an NFG of $F$ nodes. Suppose, among $F$ nodes, $d_1$ have degree 1, $d_2$ have degree 2,..., $d_D$ have degree $D$, where $D$ is the maximum degree of a node in the graph. For simplicity, we assume that 1 CPU cycle is required for computing message at any node and that time is negligible at any edge. Then the total complexity (in CPU cycles) in the graphical method can be written as \cite{Wymeersch2007}, $C_{FG}\!\!=\!\! \sum^{D}_{i=1} id_i |X|^{i}$, where $|X|$ is the cardinality of the domain of RVs. Following the conventional (explicit) method, it is required to integrate out all other variables which has a complexity equal to $C_{CN}\!=\!M{|X|}^{M}$. Thus, the NFG-SPA based approach ($O(|X|^{D})$) is computationally more efficient than the explicit method ($O(|X|^{M})$. In general $D<<M$, hence $C_{FG}\!<<\!C_{CN}$.
\\
\textit{Example:} Consider the graph of Fig. \ref{Fig6} with $K\!=\!1$, i.e. the case of $M=6$ and assume $|X|\!=\!2$. Therefore, the total complexity to find the marginals using the graphical method is 
\begin{align}\label{eq41}
C_{FG}\!=[1\!\times\!2\!\times\!\!2^1\!+\!2\!\times\!1\!\times\!2^2\!+\!3\!\times\!2\!\times\!2^3]=60 \ \text{cycles}.
\end{align}
In comparison, the complexity in the numerical method is
\begin{align}\label{eq42}
C_{CN} \!=6\times2^{6}=384 \ \text{cycles}.
\end{align}
It is interesting to note that NFG is computationally more efficient by $\frac{384}{60}\!\approx\!6.4$ times than the conventional method.
\begin{figure}[!t]
\centering
\vspace{-0.1cm}
\includegraphics[scale=0.53]{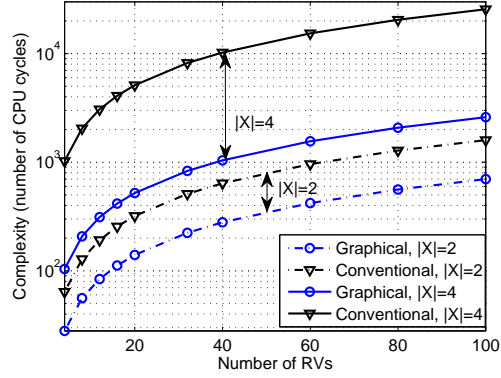}
\vspace{-0.2cm}
\caption{Computation complexity vs. number of RVs involved in computations over Nakagami-$1$ faded PU-SU and SU-FC channels.}
\label{Fig8}
\end{figure}
\par
To understand the advantage of the factor graph for the distributed detection problem addressed in this paper, Fig. \ref{Fig8} plots the computation complexity as a function of the number of RVs involved. The desired likelihood $P(y_{k}|H_i)$ is obtained by marginalizing $P(z_{k},t_{k},u_{k},v_{k},y_{k}|H_i)$ in \eqref{eq21}. Each variable of interest of the function is defined over the same $X$ discrete-sampled values from a continuous domain (i.e. uniformly quantized with $Q\!=\!|X|\!=\!4$ levels). We consider two cases as $|X|\!=\!2$, and $|X|\!=\!4$ for $K\!=\!10$. From the previous discussions and Fig. \ref{Fig6}, in this case $M=4$, $D=2$. The total complexities to find likelihoods in both conventional (\eqref{eq21}) and graphical (\eqref{eq22}) methods are computed as 

\small
\begin{equation*}
  \left.
	\begin{aligned}
   &C^{2}_{FG}\!=\! \sum^{D}_{i=1} id_i |X|^{i} \!=\!\!\left[1\! \times\! 2\! \times\! 2^{1}\!+\!2\times\! 3\! \times\! 2^{2}\right]\!\times\! 10 \!\!=\!280 \ \textit{cycles}\quad\\
   &C^{2}_{CN}\!=\! M|X|^{M} \!=\!\left[4\times 2^{4}\right] \times 10\!=\!640 \ \textit{cycles}
\end{aligned}
	\right\},
\end{equation*}
\normalsize
for $|X|\!=\!2$, and
\small
\begin{equation*}
  \left.
	\begin{aligned}
   &C^{4}_{FG}\!=\! \sum^{D}_{i=1} id_i |X|^{i} \!=\!\!\left[1\! \times\! 2 \!\times\! 4^{1}\!+\! 2 \!\times\! 3 \!\times \! 4^{2}\right]\!\times\! 10 \!\!=\!1040 \ \textit{cycles}\quad\\
   &C^{4}_{CN}\!=\! M|X|^{M} \!=\!\left[4\times 4^{4}\right] \times 10\!=\!10240 \ \textit{cycles}
\end{aligned}
	\right\},
\end{equation*}
\normalsize
for $|X|\!=\!4$, respectively. Fig. \ref{Fig8} shows that the complexity monotonically increases with the number of RVs involved (i.e. as the system becomes more complex) in both methods. It also shows that complexity increases as the quantization level $Q$ increases. However, it is interesting to note that the rate of increment in NFG-SPA setting is considerably less than that of the explicit method. It implies that NFG-SPA setting is computationally more efficient (e.g. $\frac{640}{280}\!\approx\!2.4$ times for $|X|\!=\!2$) than the conventional method (for acyclic FG). Therefore, we can state that for a large network and higher quantization level, graphical method is more effective from an implementation point of view, even for statistically independent channels. Moreover, collaborative spectrum sensing is very effective, when collaborating CRs observe independent fading \cite{Yucek2009}, which results in acyclic FG. Hence, NFG-SPA settings are suitable from implementation perspective.

\section{Conclusions}
In this paper the problem of centralized cooperative spectrum sensing over Nakagami-$m$ fading in a CRN, was addressed. We have presented a new fusion rule based on LRT, which requires only the statistical characteristics of the wireless channels between PU-SU-FC. It was compared with other suboptimal rules e.g. EGC, MRC and better performance has been observed. Furthermore, we derived a novel expression for the local probability of detection for ED over Nakagami-$m$ fading. It is shown that the proposed models perform better, over a wide range of SNR for different values of $m$, compared to the approximate complex signal representations. 
Closed-form solutions for the system-level probabilities of detection ($P_D$) and false alarm ($P_F$) were derived. Furthermore, an expression for the optimal number of cooperative SUs, needed to minimize the total error rate, is also obtained. It leads to a fast spectrum sensing algorithm, by considering only the optimal number ($l_{opt}$) of cooperative SUs instead of the total one. 
Furthermore, decision fusion based on SPA over representative factor graphs was used. It was shown that a considerable amount of gain in computational complexity can be achieved through NFG-SPA settings, even if PU-SU-FC channels are assumed to be statistically independent. 
More complex channel conditions, e.g. correlated PU-SU-FC channels, and design of fusion rules for soft decisions can be investigated as future works. 





\appendices
\section{Message Passing on Factor Graph of Fig. \ref{Fig6}}
\vspace{-0.6cm}
\small
\begin{align}\label{eq43}
&(i)M_{y_k\rightarrow P(y_k|u_k)} = 1 \ \text{;} \ (ii)M_{P({v_k})\rightarrow v_k} = P({v_k}) \nonumber \\
&(iii)M_{v_k\rightarrow P(y_k|u_k,{v_k})} = P({v_k}) \text{;}\ (v)M_{u_k \rightarrow P(u_k|t_k)}\!\!\! =\!P(y_k|u_k) \nonumber \\
&(iv)M_{P(y_k|u_k,v_k)\rightarrow u_k}\!\!\!=\!\!\!\int^{}_{v_k}\!\!\! P(y_k|u_k,v_k)P({v_k}) \ dv_k=P(y_k|u_k)\nonumber \\
&(vi)M_{P(u_k|t_k)\rightarrow t_k}\!=\!P(y_k|-1) I(t_k\!<\!\tau_k)\!+\! P(y_k|1) I(t_k\!>\!\tau_k) \nonumber \\ 
&(vii)M_{t_k\rightarrow P(t_k|H_i,{z_k})}\! =\! P(y_k|-1) I(t_k\!<\!\tau_k)\! +\! P(y_k|1) I(t_k\!>\!\tau_k) \nonumber \\
&(viii)M_{P({z_k})\rightarrow z_k} = P({z_k}) \nonumber \\
&(ix)M_{z_k\rightarrow P(t_k|H_i,{z_k})} = P({z_k}) \ \text{and} \ (x)M_{H_i \rightarrow P(t_k|H_i,{z_k})} = 1 \nonumber \\
&(xi)M_{P(t_k|H_i,{z_k})\rightarrow H_i} = \int^{}_{t_k}\!\!\int^{}_{z_k}\!\!\!\!\!\!P(t_k|H_i,{z}_k)dz_k dt_k \times(ix)\times(vii)\nonumber \\ 
&=\!\! \int^{}_{t_k}\!\!\int^{}_{z_k}\!\!\!P({z}_k)P(t_k|H_i,{z_k})\ [P(y_k|u_k=-1) I(t_k<\tau_k) \nonumber \\
   &+ P(y_k|u_k=1) I(t_k>\tau_k)]\ dz_k dt_k 
\end{align}
\normalsize
\section{Steps from (14) to (15)}
Integrating \eqref{eq12} using Laguerre polynomials \cite[(6.9.2.36)]{Erdelyi1953}, \cite{wolfram}, the \textit{cdf} of $t_k$ is obtained as

\vspace{-0.3cm}
\small
\begin{align}\label{eqA}
F_{Nak}(t_k|H_1)\!= M\sum^{\infty}_{i=0}\!\frac{{\left(\alpha_{{ps}_k}\!\right)}^{i}\!\left(m\right)_i}{i! \ (N+i)!} \gamma\left(\!N\!+\!1\!+\!i,\frac{\tau_k}{2\sigma^{2}_{w_k}}\!\right) 
\end{align}
\normalsize
In \eqref{eqA}, $(G)_i\!=\!\frac{\Gamma(G+i)}{\Gamma(G)}$ for $i\!=\!0,1,2,..\infty$ is the Pochhammer symbol \cite{Abramowitz1972}. Using \cite[(6.5.12)]{Abramowitz1972} in \eqref{eqA} and then \cite[(13.1.2)]{Abramowitz1972}, we get
\small
\begin{align}
&F_{Nak}(t_k|H_1)\!=\nonumber \\
&M \sum^{\infty}_{i=0}\!\frac{{\left(\alpha_{{ps}_k}\!\right)}^{i}\!\left(m\right)_i}{i! \ (N+i)!} \! \times \! \frac{\left(\!\!\frac{\tau_k}{2\sigma^{2}_{w_k}}\!\!\right)^{(\!N+\!1+\!i\!)} \!\!{}_1\!F_1\left(\!1;N\!+\!1\!+\!i\!+\!1;\frac{\tau_k}{2\sigma^{2}_{w_k}}\!\!\right)}{(N+1+i)\ e^{\frac{\tau_k}{2\sigma^{2}_{w_k}}}}\nonumber \\
&=\! \frac{M\left(\frac{\tau_k}{2\sigma^{2}_{w_k}}\right)^{(N+1)}}{N! \ e^{\frac{\tau_k}{2\sigma^{2}_{w_k}}}} \Phi_2\left(\!m,1;N\!+\!1;\frac{\alpha_{{ps}_k}\tau_k}{2\sigma^{2}_{w_k}},\! \frac{\tau_k}{2\sigma^{2}_{w_k}} \!\right).
\end{align}
\normalsize
where $M\!=\!{\left(\!\frac{m}{m\!+\!\bar{\rho}^{(m)}_{{ps}_k}}\!\right)}^m$ and $(a)!\left(a\!+\!1\right)_{j}\!=\!N!\left(N+1\right)_{i+j}$, when $a=N+1+i$.

\section{Derivation of Eq. (16)} 
For $m\!\!=\!\!1$, using the values of ${}_1\!F_1\left(\frac{3}{2},\frac{3}{2},a \right)$, ${}_1\!F_1\left(1,\frac{1}{2},a\right)$, $\Gamma\left(\frac{3}{2}\right)$ \cite{Abramowitz1972}
, and after some algebra, \eqref{eq11} may be simplified as (without loss of generality, index $k$ is dropped from variables $A,B,D$)
\small
\begin{align}\label{eq44}
&P^{(1)}_{|x_k|}(a|H_1)\!\!=\!\frac{\sigma_{w_k}e^{-\frac{a^{2}}{2\sigma^{2}_{w_k}}}}{A_{w}\sqrt{2\pi}}\!\left[a e^{a^{2} D_w}\sqrt{\pi D_w} Q\left(\!- \frac{\sigma_{z_k} {a}}{\sigma_{w_k}\sqrt{A_{w}}}\!\right) \!+\!1\right] \nonumber \\
&=\!\frac{ \sigma_{w_k}}{A_{w} \sqrt{2\pi}} e^{\!-\frac{{a}^2}{2\sigma^{2}_{w_k}}}\!+\! \frac{\sigma_{w_k}B_{w} a}{A_{w}} \ e^{\!-\frac{{a}^2}{2A_{w}}}\ Q\left(-B_{w} a\right).
\end{align}
\normalsize
Similarly, for $m\!=\!2$, using the values of $\Gamma\left(\frac{5}{2}\right)$, ${}_1\!F_1\left(\frac{5}{2},\frac{3}{2},a\right)$, and ${}_1\!F_1\left(2,\frac{1}{2},a \right)$ \cite{Abramowitz1972}
, and after some algebra, \eqref{eq11} becomes 
\small
\begin{align}\label{eq45}
&P^{(2)}_{|x_k|}(a|H_1)\! \nonumber \\
&=\!\frac{\sigma^{3}_{w}B_{w} a}{2{A_{w}}^{2}}\left[\!\frac{a B_{w}  e^{\!-\frac{{a}^2}{2\sigma^{2}_{w_k}}}}{\sqrt{2\pi}} \!+\!{\left(2\!+\!{B^{2}_{w} a^{2}}\!\right)} e^{\!-\frac{{a}^2}{2A_{w}}}\ Q\left(-B_{w} a\right)\!\right],
\end{align}
\normalsize
where $B_{w}\!=\!\frac{\sigma_{z_k}}{\sigma_{w_k}\sqrt{A_{w}}}$ and $A_{w}\!=\!\left(\sigma^{2}_{z_k}+\sigma^{2}_{w_k}\right)$. 

\section{Derivation of Eq. (20)}
As $\gamma\left(N,a\right)\!=\! N^{-1}a^{N}e^{-a} {}_1F_1\left(1;N+1;a \right)$ \cite[(6.5.12)]{Abramowitz1972}, and
using \cite[(13.4.3), (6.5.12)]{Abramowitz1972}, we can write
\small
\begin{align}\label{eq46}
&{}_1F_1\left(2;N+1;a \right)\!=\! (1-N) {}_1F_1\left(1;N+1;a \right) \!+\! N {}_1F_1\left(1;N;a \right) \nonumber \\
&=\! \frac{N(N-1)}{a^{N}}e^{a} \left\{a\gamma\left(N-1,a\right)\!-\!\gamma\left(N,a\right)\right\}.   
\end{align}
\normalsize
Therefore, \eqref{eq17} can be simplified as 
\small
\begin{align}\label{eq47}
&P_{(m=2)}(t_k|H_1)\!=\! \nonumber \\
&\frac{\left(\!1\!-\!\alpha_{{ps}_k}\!\right)^{2}e^{-\frac{\left(\!1\!-\!\alpha_{{ps}_k}\!\right)t_k}{2\sigma^{2}_{w_k}}}}{2\sigma^{2}_{w_k}\Gamma(N-1)\left(\!\alpha_{{ps}_k}\!\right)^{N}}\!\left[\!\frac{\alpha_{{ps}_k}t_k}{2\sigma^{2}_{w_k}}\gamma\left(\!\!N\!-\!1,\!\frac{t_k \alpha_{{ps}_k}}{2\sigma^{2}_{w_k}}\!\!\right)\!-\!\gamma\left(\!\!N,\!\frac{t_k \alpha_{{ps}_k}}{2\sigma^{2}_{w_k}}\!\!\right)\!\right].
\end{align}
\normalsize
Itegrating \eqref{eq47} over $t_k$ by parts, then, using $\int\gamma\left(N,a\right) da\!=\!a\gamma\left(N,a\right)\!+\!\Gamma\left(N+1,a\right)$ and \cite[(1.2.2.1)]{Prudnikov1992} we get the local probability of detection for $m=2$ in \eqref{eq18}.

\section{Derivation of Eq. (35)}

Consider, for all SUs $p_{d_k}$'s and $p_{f_k}$'s are same i.e. $p_{d_k}\!=\!p_d$ and $p_{f_k}\!=\!p_f$ $\forall k$. 
As $u_k\!\in\!\left\{-1,1\right\}$, $P(y_k|u_k)$ follows \eqref{eq11}. Thus, for $m\!=\!1$ $P(y_k|u_k)$ follows \eqref{eq14}, by replacing $x_k,\sigma^{2}_{z_k}$ and $\sigma^{2}_{w_k}$ with $y_k,\sigma^{2}_{v_k}$ and $\sigma^{2}_{n_k}$, respectively. Then, 
\small
\begin{equation}\label{eq48}
  P(y_k|u_k)\!=\!\frac{\sigma_{n_k}e^{-\frac{{y_k}^2}{2\sigma^{2}_{n_k}}}}{A_n \sqrt{2\pi}} \!+\! \frac{u_k\sigma_{n_k}B_n y_k}{A_n} e^{\!-\frac{{y_k}^2}{2A_n}} Q(-u_kB_n y_k).
\end{equation}
\normalsize
Therefore, likelihood under $H_1$ is written as
\small
\begin{align}\label{eq49}
	  &P_{(m=1)}(y_k|H_1)\!= P(y_k|u_k=-1)(1-p_{d})+P(y_k|u_k=1)p_{d} \nonumber \\
	&=\!\! \frac{\sigma_{n_k}}{A_{n}\sqrt{2 \pi}}\!\! \left[\!e^{-\frac{{y_k}^2}{2\sigma^{2}_{n_k}}} \!+\!\left\{p_{d}\!-\!Q(\!B_{n} y_k\!)\right\}\!\sqrt{2 \pi} B_{n} {y_k} e^{-\frac{{y_k}^2}{2A_{n}}}\!\right],
\end{align}
\normalsize
where $A_n\!\!=\!\!\left(\sigma^{2}_{v_k}\!+\!\sigma^{2}_{n_k}\right)$ and $B_n\!\!=\!\!\frac{\sigma_{v_k}}{\sigma_{n_k}\sqrt{A_n}}$. We further compute
\vspace{-0.2cm}
\small
\begin{align}\label{eq50}
&p_1\!=\!P(y_k \geq 0|H_1)\!=\! \int^{\infty}_{0}\!\! P(y_k|H_1)\ dy_k \nonumber \\
&=\! \frac{\sigma_{n_k}}{A_{n}\sqrt{2 \pi}}\left[\!\sqrt{\frac{\pi}{2}}\sigma_{n_k}\!+\!p_{d}\sqrt{2\pi}B_{n} A_{n}\!-\!\frac{\sqrt{2\pi}B_{n}}{2}\left(\!A_n-\sigma_{v_k}\sqrt{A_n}\!\right)\!\right] \nonumber \\
&=\! \frac{1}{2}\!+\!\frac{(p^{(m=1)}_{d}-\frac{1}{2})\sigma_{v_k}}{\sqrt{\sigma^{2}_{v_k}\!+\!\sigma^{2}_{n_k}}}. 
\end{align}
\normalsize
For $m=1$, by replacing $p_{d}$ with $p_{f}$ in \eqref{eq49} and \eqref{eq50} we get $P_{(m=1)}(y_k|H_0)$ and $p_0$, respectively as 
\small
\begin{align}\label{eq51}
p_0 = P(y_k \geq 0|H_0)=\frac{1}{2}\!+\!\frac{(p_{f}-\frac{1}{2})\sigma_{v_k}}{\sqrt{\sigma^{2}_{v_k}\!+\!\sigma^{2}_{n_k}}}.
\end{align}
\normalsize
Similarly, for $m\!=\!2$, $P(y_k|u_k)$ follows \eqref{eq45}. Thus, we get 

\small
\begin{align}\label{eq52}
	&P(y_k|u_k)=\frac{\sigma^{3}_{n_k} B_n y_k}{2A^{2}_n} \nonumber \\
	&\times\left[\frac{B_n y_k}{\sqrt{2\pi}}e^{\!-\frac{{y_k}^2}{2\sigma^{2}_{n_k}}} \!+\! {u_k}\left(2+B^{2}_n y^{2}_k\right) e^{\!-\frac{{y_k}^2}{2A_n}}\ Q\left(-u_k B_n y_k\right)\!\right],
		\end{align}
\normalsize
and the likelihood under $H_1$ as 
\small
\begin{align}\label{eq53}
    &P_{(m=2)}(y_k|H_1)=\frac{\sigma^{3}_{n}B_n y_k}{2\sqrt{2\pi}A^{2}_{n}} \nonumber \\
		&\times\left[\!B_n {y_k} e^{-\frac{{y^{2}_k}}{2\sigma^{2}_{n_k}}}\!\!+\!\left\{p_{d_k}\!-\!Q(B_{n} y_k)\right\}\!\sqrt{2\pi}\left(2\!+\!B^{2}_n y^{2}_k\right)e^{-\frac{{y^{2}_k}}{2A_{n}}}\!\right].		
\end{align}
\normalsize
Then, $p_1$ is computed from \eqref{eq53}, using \cite[(1.5.3.1,8)]{Prudnikov1992}, as 
\small
\begin{align}\label{eq54}
p_1\!=\! \frac{\sigma^{2}_{v_k}}{\left(\sigma^{2}_{v_k}\!+\!\sigma^{2}_{n_k}\right)^{2}}\left(\frac{\sigma^{2}_{v_k}}{2\sigma_{n_k}}+\sigma_{n_k}\right)\!+\!\frac{(p^{(m=2)}_{d}-\frac{1}{2})\sigma_{v_k}}{\sigma_{n_k}\sqrt{\sigma^{2}_{v_k}\!+\!\sigma^{2}_{n_k}}}. 
\end{align}
\normalsize
Similarly, for $m=2$, by replacing $p_{d}$ with $p_{f}$ in \eqref{eq53} and \eqref{eq54}, we get $P_{(m=2)}(y_k|H_0)$ and $p_0$, respectively. 

\section{Derivation of $p^{(m=\frac{1}{2})}_{d_k}$ and the LR for $m=\frac{1}{2}$.}
As discussed earlier, $E[{|z_k|}^2]\!\!=\!\!2m\sigma^{2}_{z_k}$ and $E[{|v_k|}^2]\!\!=\!\!2m\sigma^{2}_{v_k}$. Setting $m=\frac{1}{2}$ in \eqref{eq12}, we get 
\vspace{-0.25cm}
\small
\begin{align}\label{eq55}
&P_{(m=\frac{1}{2})}(t_k|H_1)\!=\!\frac{{t_k}^{N}e^{-\frac{t_k}{2\sigma^{2}_{w_k}}}{\left(\!\frac{\frac{1}{2}}{\frac{1}{2}+\bar{\rho}^{(\frac{1}{2})}_{{ps}_k}}\!\right)^{\frac{1}{2}}}}{\left(N\right)!\left(2 \sigma^{2}_{w_k}\right)^{N+1}} \nonumber \\
&\times\!{}_1\!F_1\left(\frac{1}{2},N+1,t_k D_{w_k} \right), 
\end{align}
\normalsize
where $\bar{\rho}^{(\frac{1}{2})}_{{ps}_k}\!=\!\frac{\sigma^{2}_{z_k}}{2\sigma^{2}_{w_k}}$ is the average SNR of $PU-SU_k$ link. Now, integrating \eqref{eq55} over $t_k$ and using \cite[(6.5.12)]{Abramowitz1972}, the probability of detection for $m=\frac{1}{2}$ may be computed as
\small
\begin{align}\label{eq56}
p^{(m=\frac{1}{2})}_{d_k}\!
=\!1-\!(2A_w)^{N}\frac{\gamma\left(\! N,\! \frac{\tau_k}{2A_w}\!\right)}{\Gamma(N)}.
\end{align}
\normalsize
As, $u_k\!\in\!\left\{-1,1\right\}$, $P(y_k|u_k)$ is obtained by replacing $x_k$, $\sigma^{2}_{w_k}$, and $\sigma^{2}_{z_k}$ with $y_k$, $\sigma^{2}_{n_k}$, and $\sigma^{2}_{v_k}$, respectively in \eqref{eq11} as  
\vspace{-0.2cm}
\small
\begin{align}\label{eq57}
P(y_k|u_k)&=\!\frac{1}{\sqrt{2\pi\sigma^{2}_{n_k}}}{\left(\frac{\sigma^{2}_{n_k}}{\sigma^{2}_{v_k}+\sigma^{2}_{n_k}}\right)}^{\frac{1}{2}} e^{-\frac{{y_k}^2}{2(\sigma^{2}_{v_k}+\sigma^{2}_{n_k})}}\nonumber \\
&\left[1+\sqrt{\frac{\sigma^{2}_{n_k}}{2}}erf\left(\frac{y_k u_k}{\sqrt{2\sigma^{2}_{n_k}(\sigma^{2}_{v_k}+\sigma^{2}_{n_k})}}\right)\right],
\end{align}
\normalsize
where $A_n\!=\!(\sigma^{2}_{v_k}+\sigma^{2}_{n_k})\!=\!\sigma^{2}_{n_k}(1+2\bar{\rho}^{(\frac{1}{2})}_{{sf}_k})$ and average SNR of $SU_k-FC$ link is defined as $\bar{\rho}^{(m\!=\!\frac{1}{2})}_{{sf}_k}\!=\!\frac{\sigma^{2}_{v_k}}{2\sigma^{2}_{n_k}}$. Now, LRT statistic may be written as \eqref{eq58}. 
\begin{figure*}
\vspace{-0.5cm}
\small
\begin{align}\label{eq58}
L(y^{K}_{1}) \!\!
=\!\prod^{K}_{k=1}\!\!\frac{\left(1-p^{(m=\frac{1}{2})}_{d_k}\right)\left[1\!-\!\sqrt{\frac{\sigma^{2}_{n_k}}{2}}erf\left(\frac{y_k}{\sqrt{2\sigma^{4}_{n_k}(1+2\bar{\rho}_{{sf}_k})}}\right)\right]\!+\!p^{(m=\frac{1}{2})}_{d_k} \left[1\!+\!\sqrt{\frac{\sigma^{2}_{n_k}}{2}}erf\left(\frac{y_k}{\sqrt{2\sigma^{4}_{n_k}(1+2\bar{\rho}_{{sf}_k})}}\right)\right]}{\left(1-p_{{f_k}}\right)\left[1\!-\!\sqrt{\frac{\sigma^{2}_{n_k}}{2}}erf\left(\frac{y_k}{\sqrt{2\sigma^{4}_{n_k}(1+2\bar{\rho}_{{sf}_k})}}\right)\right]\!+p_{{f_k}} \left[1\!+\!\sqrt{\frac{\sigma^{2}_{n_k}}{2}}erf\left(\frac{y_k}{\sqrt{2\sigma^{4}_{n_k}(1+2\bar{\rho}_{{sf}_k})}}\right)\right]}.
\end{align}
\normalsize
\vspace{-0.5cm}
\end{figure*}
For both $m\!=\!\frac{1}{2}$ and $2$, LRT thresholds ($\lambda$) are obtained by (numerically) solving $\int^{\infty}_{\lambda}\!P(L|H_0) \ dL\!=\!P_F$, for given probabilities of false alarm. 
\section{Derivation of Eq. (21) and Eq. (22)}
\textit{\underline{Case-I: $\frac{1}{2}\leq m<1$}:}
In this case, the complex Nakagami-$m$ envelope are expressed as Hoyt approximation \cite[(61)]{Mallik2010}. The real and imaginary parts are \mbox{\small{$\mathcal{N}\left(0,\frac{\Omega_{z_k}(1+b)}{2}\right)$}} and \mbox{\small{$\mathcal{N}\left(0,\frac{\Omega_{z_k}(1-b)}{2}\right)$}}, respectively, where $b\!\!=\!\!\sqrt{\frac{1-m}{m}}$. Therefore, \mbox{\small{$z_k(n)\sim \mathcal{CN}\left(0,\frac{\Omega_{z_k}\left(1+b\right)}{2}+\frac{\Omega_{z_k}\left(1-b\right)}{2}\right)$}}. Now, as \mbox{\small{$w_k(n)\sim \mathcal{CN}\left(0,2\sigma^{2}_{w_k}\right)$}}, we may write \newline \mbox{\small{$x_k(n)\sim \mathcal{CN}\left(0,\left(\frac{\Omega_{z_k}\left(1+b\right)}{2}+\frac{\Omega_{z_k}\left(1-b\right)}{2}+2\sigma^{2}_{w_k}\right)\right)$}}. The test statistic $t_k$ is the sum of $N$ squares of \textit{i.i.d.} \mbox{\small{$\mathcal{N}\left(0,\left[\frac{\Omega_{z_k}\left(1+b\right)}{2}\!+\!\sigma^{2}_{w_k}\right]\right)$}} and another $N$ squares of \textit{i.i.d.} \mbox{\small{$\mathcal{N}\left(0,\left[\frac{\Omega_{z_k}\left(1-b\right)}{2}\!+\!\sigma^{2}_{w_k}\right]\right)$}} real Gaussian RVs with each having zero mean. 
Thus, the \textit{pdf} of $t_k$ under $H_1$ is the sum of two central Chi-square distributions with each having $N$ degrees of freedom (DOF) and written as $P_{(\frac{1}{2}\leq m<1)}(t_k|H_1)\!=$

\vspace{-0.2cm}
\small
\begin{align}\label{eq59}
\frac{{t_k}^{\frac{N}{2}-1} e^{-\frac{t_k}{\left(\Omega_{z_k}\left(1+b\right)+2\sigma^{2}_{w_k}\right)}}}{{\Gamma\left(\frac{N}{2}\right)}{\left(\Omega_{z_k}\left(1+b\right)+2\sigma^{2}_{w_k}\right)}^{\frac{N}{2}}}\!+\! \frac{{t_k}^{\frac{N}{2}-1} e^{-\frac{t_k}{\left(\Omega_{z_k}\left(1-b\right)+2\sigma^{2}_{w_k}\right)}}}{{\Gamma\left(\frac{N}{2}\right)}{\left(\Omega_{z_k}\left(1-b\right)+2\sigma^{2}_{w_k}\right)}^{\frac{N}{2}}}.
\end{align}
\normalsize
Hence, for $\frac{1}{2}\leq m<1$, the corresponding probability of detection at each $SU_k$ is obtained by integrating \eqref{eq59} as
\vspace{-0.2cm}
\small
\begin{align}\label{eq60}
p^{\left(\frac{1}{2}\!\leq\! m\!<\!1\right)}_{d_k} 
\!\!=\!1-\!\frac{\gamma\left(\!\!\frac{N}{2},\frac{\tau_k}{\left(\Omega_{z_k}\left(1+b\right)+2\sigma^{2}_{w_k}\right)}\!\!\right)}{\Gamma\left(\frac{N}{2}\right)}\!-\!\frac{\gamma\left(\!\!\frac{N}{2},\frac{\tau_k}{\left(\Omega_{z_k}\left(1-b\right)+2\sigma^{2}_{w_k}\!\right)}\!\!\right)}{\Gamma\left(\frac{N}{2}\right)}. 
\end{align}
\normalsize
\par
\textit{\underline{Case-II: $m=1$}:}
This is the case of Rayleigh distribution where \mbox{\small{$z_k(n)\sim \mathcal{CN}\left(0,2\sigma^{2}_{z_k}\right)$}} \cite{Atapattu2014}. Then, the signal is written as \mbox{\small{$x_k(n)\sim \mathcal{CN}\left(0,2\left(\sigma^{2}_{z_k}+\sigma^{2}_{w_k}\right)\right)$}}. Therefore, the \textit{pdf} of test statistic $t_k$ under $H_1$ follows central Chi-square distribution with $2N$ DOF. The probability of detection at $SU_k$ is computed by integrating $P_{(m=1)}(t_k|H_1)$ as \cite{Zarrin2008}
 
\vspace{-0.3cm}
\small
\begin{align}\label{eq61}
p^{(m=1)}_{d_k}\! 
=\!1-\!\frac{\gamma\left(\!N,\frac{\tau_k}{2\left(\sigma^{2}_{z_k}+\sigma^{2}_{w_k}\right)}\!\right)}{\Gamma\left(N\right)}. 
\end{align}
\normalsize
\par
%
 %
\par
\textit{\underline{Case-III: $m>1$}:}
In this case, the complex Nakagami-$m$ envelope are expressed as Rician approximation \cite[(59)]{Mallik2010}. Here, the real and imaginary parts are \mbox{\small{$\mathcal{N}\left(\mu_{I_{z_k}},\frac{\Omega_{s_k}}{2}\right)$}} and \mbox{\small{$\mathcal{N}\left(\mu_{Q_{z_k}},\frac{\Omega_{s_k}}{2}\right)$}} distributed, respectively. Therefore, \mbox{\small{$z_k(n)\sim \mathcal{CN}\left(\mu_{I_{z_k}}\!+\!j\mu_{Q_{z_k}}, \Omega_{s_k}\right)$}}. 
Now, as \mbox{\small{$w_k(n)\sim \mathcal{CN}\left(0,2\sigma^{2}_{w_k}\right)$}}, we may write \mbox{\small{$x_k(n)\sim \mathcal{CN}\left(\mu_{I_{z_k}}+j\mu_{Q_{z_k}},\left(\Omega_{s_k}+2\sigma^{2}_{w_k}\right)\right)$}}. Then, the test statistic ($t_k$) is a sum of $2N$ squares of independent and non-identically ditributed Gaussian RVs with each having non-zero mean. Therefore, the \textit{pdf} of $t_k$ under $H_1$ follows non-central Chi-square distribution with $2N$ degrees of freedom and non-centrality parameter \mbox{\small{$\mu_{z_k}\!=\!\sum^{N}_{n=1}\left[\frac{\mu^{2}_{I_{z_k}}}{\left(\frac{\Omega_{s_k}}{2}+\sigma^{2}_{w_k}\right)}+\frac{\mu^{2}_{Q_{z_k}}}{\left(\frac{\Omega_{s_k}}{2}+\sigma^{2}_{w_k}\right)}\right]\!=\!\frac{2N\bar{\rho}_{{ps}_k}\sqrt{\frac{m-1}{m}}}{\bar{\rho}_{{ps}_k}\left(1-\sqrt{\frac{m-1}{m}}\right)+1}$}}. With help of \cite{Abramowitz1972}, it may be written as 

\vspace{-0.2cm}
\small
\begin{align}\label{eq62}
P_{(m>1)}(t_k|H_1)\!=\!\frac{\left(\frac{2t_k}{\Omega_{x_k}}\right)^{\frac{N-1}{2}}}{\left(\Omega_{x_k}\right)^{\frac{N-1}{2}}e^{\frac{1}{2}\left(\frac{2t_k}{\Omega_{x_k}}+\mu_{z_k}\right)}} I_{N-1}\left(\sqrt{\frac{\mu_{z_k}2t_k}{\Omega_{x_k}}}\right),
\end{align}
\normalsize
where $0\!\leq \!t_k\!\leq\!\infty$, $\Omega_{x_k}\!=\!\Omega_{s_k}+2\sigma^{2}_{w_k}$, and $I_\nu(.)$ is the modified Bessel function of the first kind of order $\nu$. The probability of detection at $SU_k$ is given as \cite[(2.13)]{Atapattu2014} 
 
\vspace{-0.2cm}
\small
\begin{align}\label{eq63}
p^{(m>1)}_{d_k}\!=\!Q_{N}\left(\sqrt{\mu_{z_k}},\sqrt{\frac{2\tau_k}{\Omega_{s_k}+2\sigma^{2}_{w_k}}}\right).
\end{align}
\normalsize
%

\ifCLASSOPTIONcaptionsoff
  \newpage
\fi



%

%
%

\bibliographystyle{IEEEtran}
\bibliography{IEEEabrv,graphicalModel}

\end{document}